\documentclass[prb,twocolumn,showpacs,amsmath,amssymb,superscriptaddress,bibnotes,floatfix]{revtex4-2}
\usepackage[dvipsnames]{xcolor}
\usepackage[T1]{fontenc}
\usepackage{amsmath}
\usepackage{braket}
\usepackage{simpler-wick}
\usepackage{tikz}

\usepackage{graphicx}
\usepackage{hyperref}
\usepackage{placeins}
\usepackage{bbold}
\usepackage{comment}

\renewcommand{\d}{\mathrm{d}}

\newcommand{\cc}[1]{{#1}^{\dagger}}
\newcommand{\dotd}{\hat d}
\newcommand{\dotc}{\hat d^{\dagger}}

\newcommand{\destroy}{\hat c}
\newcommand{\create}{\hat c^{\dagger}}

\renewcommand{\v}[1]{\ensuremath{\mathbf{#1}}}

\usepackage{tikz}
\usetikzlibrary{positioning,shapes.arrows}

\begin{document}
\preprint{APS/123-QED}

\title{Many-body correlations in Floquet steady-states: Frequency-resolved renormalization group of the driven Anderson impurity}

\author{Jan-Niklas Herre}%
 \email{jan.herre@rwth-aachen.de}
 \affiliation{Institute for Theory of Statistical Physics, RWTH Aachen University and JARA—Fundamentals of Future Information
Technology, 52056, Aachen, Germany}
\author{Christoph Karrasch}
 \affiliation{Technische Universität Braunschweig, Institut für Mathematische Physik, Mendelssohnstraße 3, 38106 Braunschweig, Germany}
\author{Dante M.~Kennes}
 \affiliation{Institute for Theory of Statistical Physics, RWTH Aachen University and JARA—Fundamentals of Future Information
Technology, 52056, Aachen, Germany}
 \affiliation{Max Planck Institute for the Structure and Dynamics of Matter,
Center for Free Electron Laser Science, Luruper Chaussee 149, 22761 Hamburg, Germany}
\date{\today}

\begin{abstract}
We introduce a functional renormalization group framework formulated directly in the Floquet steady-state that systematically incorporates frequency-dependent interaction effects. By retaining the frequency structure of the two-particle vertex up to second order in interaction strength, our approach provides controlled access to dynamical response functions and nonequilibrium transport in driven, interacting systems. Using the periodically driven single-impurity Anderson model as a paradigmatic example, we benchmark our results against state-of-the-art Floquet Green’s function methods and find quantitative agreement for finite-frequency observables up to intermediate interaction strengths. Remarkably, we also show that static properties are often captured reliably by much simpler approximations, suggesting practical pathways for modeling driven quantum materials. Finally, we demonstrate that although periodic driving of the dot strongly broadens the Kondo resonance through inelastic scattering, it leaves the many-body Kondo cloud largely intact. This robustness suppresses Floquet replicas of the Kondo peak and leads to a partial persistence of Kondo pinning, highlighting the resilience of emergent many-body correlations under local periodic driving.
\end{abstract}

\maketitle

\section{Introduction}
In recent years, nonequilibrium control of quantum materials has advanced rapidly, enabling the exploration of novel physical phenomena and the ultrafast manipulation of quantum properties \cite{Basov_2017, Torre_2021, Bao_2021}. At the same time, the complexity of nonequilibrium quantum many-body dynamics has severely constrained theoretical progress, which often relies either on simplified toy models or on uncontrolled approximations whose validity in the presence of strong correlations is difficult to assess. Developing theoretical frameworks that can reliably describe interacting systems out of equilibrium therefore remains a central challenge, with direct relevance for both fundamental understanding and the design of quantum devices. \\
Floquet engineering, where time-periodic fields dress the electronic degrees of freedom and qualitatively modify their behavior, is one of the most prominent examples of nonequilibrium control. Early successes include the prediction and observation of photo-induced topological phenomena, such as the anomalous Hall effect in graphene \cite{Oka_2009, McIver_2019}, which established the field of Floquet-engineered topological phases \cite{Oka_2019, Sato_2019, Topp_2019, Wang_2013, Usaj_2014, Rodriguez_2021, Day_2024, Merboldt_2025}. 
More recently, alongside experimental advances~\cite{Boschini_2024}, increased attention has been focused on the interplay between Floquet-modified single-particle properties and many-body interactions, which enables the control of collective phenomena such as spin or charge order \cite{Mentink_2015, Claassen_2017, Golez_2019, Sarkar_2024, Quito_2021, Stojchevska_2014, Maklar_2023}, excitonic states~\cite{Baykusheva_2026} and superconductivity~\cite{Fausti_2011, Liu_2020, Sentef_2016, Kennes_2017, Kennes2018, Kennes_2019, Decker_2020, Buzzi_2020, Takahashi_2025, Sheikhan_2020, Kennes_2019, Hart_2019, Wang_2021, Kumar_2021, Dehghani_2021, Li_2017, Ning_2024, Sheikhan_2022, Tsuji_2015, Dasari_2018, Sentef_2017, Knap_2016, Murakami_2017, Ojeda_2021, Kuhn_2024}. \\
Understanding how interactions reshape Floquet steady-states is therefore a prerequisite for controlled Floquet engineering in correlated quantum matter.
Driven quantum dots constitute a particularly useful setting in this context. On the one hand, they serve as fundamental building blocks for quantum information processing and nanoelectronic devices \cite{Hanson_2007, Ludovico_2016, Haughian_2017, Eissing_2016, Eissing_2016a}. On the other hand, their reduced dimensionality allows for controlled theoretical treatments that can disentangle local decoherence effects from genuinely many-body correlations. As such, quantum impurity systems provide an ideal laboratory for developing and benchmarking theoretical approaches to driven, interacting systems.\\
A wide range of numerical methods has been developed to address nonequilibrium quantum dynamics, including quasi–numerically exact approaches such as the density matrix renormalization group (DMRG) \cite{White_1992a, White_2004, Schollw_ck_2011} and Green’s function–based techniques, such as the solution of the Kadanoff–Baym equations \cite{Schl_nzen_2019, Schuler_2020}, nonequilibrium dynamical mean-field theory (DMFT) \cite{Aoki_2014}, and the functional renormalization group (FRG) \cite{Karrasch_2008, Jakobs_2010a, Jakobs_2010b, Kennes_2012}. A common limitation of real-time approaches is the restriction to finite simulation times, which complicates access to nonequilibrium (Floquet-synchronized) steady-states. This difficulty can be circumvented if the problem is formulated directly in the infinite-time limit, as is possible within the framework of Floquet Green’s functions \cite{Ono2018, Ono2019}. This approach provides systematic access to Floquet steady-states and forms a natural starting point for incorporating interaction effects.\\
In this work, we employ the functional renormalization group to treat interactions in Floquet steady-states. In equilibrium, FRG has proven highly effective for analyzing correlated systems and identifying dominant electronic instabilities \cite{Klebl_2020, Profe_2022, Hille_2020}. In nonequilibrium settings, and particularly for periodically driven systems, FRG developments have so far largely focused on self-energy–based truncation schemes and on transport properties of quantum impurity models. Extending FRG to reliably capture interaction effects beyond static self-energy renormalization is therefore an important step toward applying the method to driven low-dimensional materials.
Here, we formulate a Floquet functional renormalization group scheme that operates directly in the steady-state and retains the frequency dependence of the two-particle vertex up to second order in the interaction. By focusing on channel-decomposed ladder diagrams, our approach effectively provides a renormalization group improvement over the random-phase approximation while remaining computationally tractable. We benchmark the method against bare second-order perturbation theory and the self-consistent GW approximation \cite{Schl_nzen_2019}, which represent alternative diagrammatic resummations in the weak-to-intermediate coupling regime. In addition, we assess simplified static approximations to the vertex flow that neglect frequency dependence, with the goal of identifying regimes where lightweight yet reliable descriptions are possible.\\
As a concrete application and benchmark, we study the periodically driven single-impurity Anderson model (SIAM). While serving as a controlled testbed for methodological developments, the SIAM also allows us to address physically relevant questions concerning the fate of many-body correlations under periodic driving. In particular, we analyze Floquet-engineered transport and the behavior of Kondo correlations in the driven steady-state~\cite{Hewson_1993, Pustilnik_2004, Kaminski_1999, Kaminski_2000, Latta_2011, Nghiem_2017, Kanasz_2018, Bruhat_2018, Krivenko_2019, Bruch_2022}, highlighting the distinct roles of local decoherence and nonlocal many-body correlations.

The paper is set up as follows: 
In Section~\ref{section: model} we briefly introduce the model to then continue to detail the method in Section~\ref{section: Method}. Here, we start by generally introducing the Floquet-Keldysh ansatz in Sections~\ref{section: Keldysh}-\ref{section: reservoirs} and then derive the FRG flow equations in Section~\ref{section: FRG}. In Section~\ref{sec:alternative_approaches} we briefly discuss the alternative approaches that we use for benchmarking and, in Section~\ref{section: Numerical implementation}, we give details about the numerical implementation. In Section~\ref{section: Results} we present the results of the FRG simulations, starting with a comparison of the different approximation schemes, including the static approximation in Sections~\ref{section: renormalized parameters} and~\ref{section: spectral}. Then, we study the Floquet engineered physics of the SIAM by first looking at the Floquet induced modifications of the Kondo resonance in Section~\ref{section: Kondo} and, next, investigating the effect of many-body correlations on the Floquet engineered transport through the dot in Section~\ref{sec: photo-induced tunneling}. We conclude with a discussion in Section~\ref{section: Discussion}.

\section{The Model}\label{section: model}
We consider the single impurity Anderson model (SIAM)~\cite{Anderson_1961, Krishna_1980a, Krishna_1980b, Hewson_1993, Peters_2006, Heidrich_2009, Jakobs_2010a, Ge2024} at zero magnetic field. The time-dependent Hamiltonian is written as 
\begin{align}
    \hat H (t) = \hat H^0_{\mathrm{dot}}(t) + \hat V_{\mathrm{dot}} + \sum_{\alpha} \hat H^{(\alpha)}_{\mathrm{coup}} + \hat H^{(\alpha)}_{\mathrm{res}} \;,
 \end{align}
 where the dot itself is modeled by a time-dependent spin-$1/2$ energy level
 \begin{equation}
     \hat H^0_{\mathrm{dot}}(t) = \sum_{\sigma} \varepsilon_{\sigma}(t)\dotc_{\sigma}\dotd_{\sigma} \;,
 \end{equation}
 and an on-site repulsive interaction
 \begin{equation}
     \hat V_{\mathrm{dot}} = U\dotc_{\uparrow}\dotd_{\uparrow}\dotc_{\downarrow}\dotd_{\downarrow} \;, \label{eq:dot_interaction}
 \end{equation}
 with $\sigma=\uparrow,\downarrow$ the electronic spin. The dot Hamiltonian explicitly depends on time $t$, varying periodically with period $T$. We incorporate a shift into the single-particle levels of $\varepsilon_\sigma(t)\to\varepsilon_\sigma(t) - U/2$ to move the particle-hole symmetric case to $\mu=0$. 
 The dot is coupled to $\alpha$ metallic reservoirs via 
 \begin{equation}
      \hat H^{(\alpha)}_{\mathrm{coup}} = \sum_{k_{\alpha},\sigma} t_{k_{\alpha}} \dotd_{\sigma} \hat c_{k_{\alpha}\sigma} + \mathrm{h.c.}
 \end{equation}
 and the reservoirs themselves are modeled by
 \begin{equation}
      \hat H^{(\alpha)}_{\mathrm{res}} = \sum_{k_{\alpha},\sigma} \varepsilon_{k_{\alpha}} \cc{\hat c}_{k_{\alpha}\sigma} \hat c_{k_{\alpha}\sigma} \;.
 \end{equation}
 At equilibrium, the SIAM provides a minimal but microscopically well-founded description of a localized, interacting electronic level hybridized with a noninteracting conduction bath. It captures the crossover from weakly correlated behavior to the local-moment regime and the emergence of Kondo screening at low energies, which manifests in the formation of a many-body singlet and a dynamically generated low-energy scale, the Kondo temperature. Owing to its conceptual simplicity, nonperturbative physics, and direct relevance to quantum dots and DMFT impurity solvers, the SIAM serves as a canonical framework for studying strong correlation effects in quantum impurities. This provides a fundamental starting point for investigations into Floquet engineering of correlated quantum systems.

\section{Method} \label{section: Method}
\subsection{Keldysh formalism} \label{section: Keldysh}
Within the Keldysh formalism \cite{Keldysh_1964, Rammer_1986, Jakobs_2010, Kamenev_2011} we set up time-dependent Green's functions on the Keldysh contour $\mathcal{C}$ in terms of fermionic operators $\hat c^{(\dagger)}_{\sigma}$ where $\sigma$ can denote any single particle degree of freedom (in this work only spin) as  
\begin{equation}
    G^{j|j'}_{\sigma|\sigma'}(t,t') = -i\langle \mathcal{T}_{\mathcal{C}}\destroy_{\sigma}^j(t) \hat c_{\sigma'}^{\dagger j'}(t') \rangle \;,
\end{equation}
where $j=\pm$ denotes the additional Keldysh contour index determining on which side of the contour the operator is applied. The expectation value of a given operator $\hat A(t)$ is defined with respect to the initial density matrix of the noninteracting system coupled to the reservoirs
\begin{equation}
    \langle \hat A\rangle (t) = \mathrm{Tr} \{\rho_0 \hat A(t)\}\;,
\end{equation}
with
\begin{equation}
    \rho_0 \equiv \rho(t_0)= \rho_{\mathrm {dot}} \otimes \bigotimes_{\alpha} \rho_{\mathrm{res}}^{(\alpha)},
\end{equation}
and the reservoirs modeled in thermal equilibrium by
\begin{equation}
    \rho_{\mathrm{res}}^{(\alpha)} = e^{-(\hat H^{(\alpha)}_{\mathrm{res}} - \mu_{\alpha}N_{\alpha})/T_{\alpha}}  /\mathrm{Tr} \{e^{-(\hat H^{(\alpha)}_{\mathrm{res}} - \mu_{\alpha}N_{\alpha})/T_{\alpha}} \}\;.
\end{equation}
Performing the Keldysh rotation, we end up with \cite{Jakobs_2010}
\begin{align}
    G_{\sigma|\sigma'}^{\mathrm R}(t,t') &= -i\Theta(t-t') \langle \{\destroy_{\sigma}(t), \create_{\sigma'}(t')]\}\rangle\;,\\
    G_{\sigma|\sigma'}^{\mathrm A}(t,t') &= i\Theta(t'-t) \langle \{\destroy_{\sigma}(t), \create_{\sigma'}(t')\}\rangle\;,\\
    G_{\sigma|\sigma'}^{\mathrm K}(t,t') &= -i\langle [\destroy_{\sigma}(t), \create_{\sigma'}(t')]\rangle\;,
\end{align}
which can be written in matrix form as
\begin{equation}
    G^{1|1'} (t,t') \equiv G_{\sigma|\sigma'}^{k|k'}(t,t') = \begin{pmatrix}
         0 & G_{\sigma|\sigma'}^{\mathrm A}(t,t')\\
         G_{\sigma|\sigma'}^{\mathrm R}(t,t') & G_{\sigma|\sigma'}^{\mathrm K}(t,t')
     \end{pmatrix}\;,
\end{equation}
where we introduced the multi-index $1=(k,\sigma)$ to combine the Keldysh index $k\in\{1,2\}$ and the remaining degrees of freedom. The frequency-dependent FRG equations are written in terms of anti-symmetrized two-particle vertex functions $\Gamma^{1'2'|12}(t_1',t_2'|t_1,t_2)$. A bare (instantaneous) interaction vertex that describes the interaction in Eq.~\eqref{eq:dot_interaction} is given by
\begin{align}
    \Gamma_{0}^{1'2'|12} (t_1',t_2'|t_1,t_2)&= \delta(t_1'=t_2'=t_1=t_2) (\Gamma_0)_{\sigma_1'\sigma_2'|\sigma_1\sigma_2} \nonumber\\
    &\times\begin{cases}
        \frac{1}{2} \;, \;\;\;\sum_i k_i \text{ odd,}\\
        0\;\;\;\;\;\;\;\text{else,}
    \end{cases}
\end{align}
with \begin{equation}
    (\Gamma_0)_{\sigma_1'\sigma_2'|\sigma_1\sigma_2} = U\delta_{\sigma_1\bar{\sigma}'_2}(\delta_{\sigma_1'\sigma_2}\delta_{\sigma_2'\sigma_1} - \delta_{\sigma_1'\sigma_1}\delta_{\sigma_2'\sigma_2})\;.
\end{equation}
\subsection{Floquet Green's and vertex functions}\label{section: Floquet Green}
To consider the synchronized steady-state of a periodically driven system, whose Green's function fulfills
\begin{equation}
   G^{1|1'} (t,t') = G^{1|1'} (t+T,t'+T) \;,
\end{equation}
we perform a double Fourier transform following the definition introduced in Refs.~\cite{Ono2018,Ono2019},
\begin{align}
    &G^{1|1'}(t,t') =  \sum_{n=-\infty}^{\infty} \int_{-\infty}^{\infty}\frac{\d \omega}{2\pi} e^{-in\Omega \overline{t} -i(\omega + \frac{n}{2}\Omega) \tau}G^{1|1'}_{n0}(\omega) \;,\\
    &G^{1|1'}_{nn'}(\omega)  \nonumber\\
    &=\frac{1}{T}\int_{0}^{T}  \d \overline{t} \int_{-\infty}^{\infty}\d \tau e^{i(\omega + n\Omega) t -i(\omega + n'\Omega) t'}G^{1|1'}(t, t')\;. \label{eq:di_floquet_Greens}
\end{align}
Here, we introduced the central time $\overline{t}=\frac{1}{2}(t+t')$ and the relative time $\tau=t-t'$. We can conveniently rewrite the Floquet Green's functions as matrices in orbital-Floquet space $(\hat G^{k|k'})_{(\sigma n)(\sigma'n')}=G^{k|k'}_{\sigma|\sigma';nn'}$, which obey the symmetry relation
\begin{equation}
    \hat G^{k|k'}(\omega)  = - (-1)^{k+k'}\left[\hat G^{k'|k}\right]^{\dagger}(\omega)  \;.
\end{equation}
The double index introduces a non-physical redundancy, so that the Green's function needs to be restricted to the ``fundamental domain'' $F=[-\Omega/2,\Omega/2)$ \cite{Rentrop2014}.
Additionally, we can define a single-index notation as
\begin{align}
    G^{1|1'}_{n}(\omega) &=  \frac{1}{T}\int_{0}^{T}\d \overline{t} \int_{-\infty}^{\infty}\d \tau e^{i\omega \tau +in\Omega \overline{t}}G^{1|1'}(t, t')\;, \label{eq:si_floquet_Greens}
\end{align}
which allows for easier numerical convolution. We can translate between the two formulations as 
\begin{align}
    G^{1|1'}_{nn'}(\omega) &= G^{1|1'}_{n-n'}\left(\omega + \frac{n+n'}{2}\Omega\right)\;,\label{eq:di_to_si}\\
    G^{1|1'}_{n}(\omega) &= G^{1|1'}_{n+n',n'}\left(\omega - \frac{2n+n'}{2}\Omega\right)\;,\label{eq:si_to_di}
\end{align}
where in the second line $n'$ has to be chosen so that it lies in the fundamental domain $(\omega - \frac{2n+n'}{2}\Omega) \in F$. Enforcing that repeated translations leave the functions invariant defines $n'$ without ambiguity. The two-particle interaction can generally only be transformed using the single-index variant~\cite{Rentrop2014}, leading to
\begin{align}
    &\Gamma^{1'2'|12}(t_1',t_2'|t_1,t_2) \nonumber\\
    &=\frac{1}{(2\pi)^3}\sum_{n=-\infty}^{\infty}\int\d {\omega} \int\d \nu \int\d \nu'\nonumber\\ 
     &  \times e^{-i\omega \tau_{\omega} - i\nu\tau_{\nu} - i\nu'\tau_{\nu'} - in\Omega \overline{t}}\Gamma^{1'2'|12}_{n}(\omega,\nu,\nu')\;, \\
    &\Gamma^{1'2'|12}_{n}(\omega,\nu,\nu') \nonumber\\ 
    &=\frac{1}{T}\int_{0}^{T}\d \overline{t} \int\d \tau_{\omega} \int\d \tau_{\nu} \int\d \tau_{\nu'}\nonumber\\ 
   &  \times e^{i\omega \tau_{\omega} + i\nu\tau_{\nu} + i\nu'\tau_{\nu'} +in\Omega \overline{t}}\Gamma^{1'2'|12}(t_1',t_2'|t_1,t_2)\;, \label{eq:si_floquet_vertex}
\end{align}
with generalized central and relative times
\begin{align}
    \bar{t} &= (t_1' + t_2' + t_1 + t_2)/4\;,\\
    \tau_{\omega} &= t_1' + t_2' - t_1 - t_2\;,\\
    \tau_{\nu} &= -t_1' + t_2' + t_1 - t_2\;,\\
    \tau_{\nu'} &= t_1' - t_2' + t_1 - t_2\;.
\end{align}
In the following, we will routinely switch between the Floquet matrix form (double index), denoted by hats, and the Floquet coefficient form (single index), which always carries a subscript Floquet index $n$.
\subsection{Floquet Dyson equation}
Any set of Keldysh diagrams can be brought into Floquet form by starting from the real-time formulation and applying Eqs.~\eqref{eq:di_floquet_Greens}-\eqref{eq:si_floquet_vertex} and their generalizations. For example, the Dyson equation in real time is given by 
\begin{align}
    G^{1|1'}(t,t') &= g^{1|1'}(t,t')\nonumber\\
    &+\int \d t_1' \d t_2 \;g^{1|2'}(t,t_1)\Sigma^{2'|2}(t_1',t_2)G^{2|1'}(t_2,t') \;,
\end{align}
with the noninteracting Green's function $g^{1|1'}(t,t')$ and the Keldysh self-energy
\begin{equation}
    \Sigma^{1'|1}(t',t) = \begin{pmatrix}
        \Sigma_{\sigma'|\sigma}^{\mathrm K} & \Sigma_{\sigma'|\sigma}^{\mathrm R}\\
        \Sigma_{\sigma'|\sigma}^{\mathrm A} & 0\\
    \end{pmatrix}\;.
\end{equation}
Due to the Keldysh rotation, the Dyson equation decouples in the Keldysh sector, and we can rewrite the retarded component in the Floquet-Keldysh formalism as a batched matrix equation 
\begin{equation}
\hat G^{\mathrm R}(\omega) = \left[\left(\hat g^{\mathrm R}\right)^{-1}(\omega) - \hat \Sigma^{\mathrm R}(\omega)\right]^{-1} \;.
\end{equation}
The advanced and Keldysh components follow as 
\begin{align}
    \hat G^{\mathrm A}(\omega) &= \left[\hat G^{\mathrm R}\right]^{\dagger}(\omega)\;,\\
    \hat G^{\mathrm K}(\omega) &= \left[\hat G^{\mathrm R}\hat \Sigma^{\mathrm K}\hat G^{\mathrm A}\right](\omega)\;.
\end{align}
\subsection{Integrating out the reservoirs}\label{section: reservoirs}
We can use the Floquet-Dyson equation to integrate out the reservoir degrees of freedom.
In the wide-band limit a time-independent reservoir $\alpha$ that is in thermal equilibrium can be modeled by the following self-energies,
 \begin{align}
    \hat \Sigma^{\mathrm{R/A}}_{\alpha} (\omega) &= \mp \frac{i}{2}\gamma_{\alpha}\mathbb{1} \;,\\ 
    \hat \Sigma^{\mathrm{K}}_{\alpha} (\omega) &= -i\gamma_{\alpha}(\mathbb{1}-2\hat f_{\alpha}(\omega)) \;.
\end{align} 
Here we introduced the momentum independent hybridization strength $\gamma_\alpha = 2\pi |t_{k_{\alpha}}|^2\rho^{(\alpha)}_{\mathrm{res}}$ and $\hat f_{\alpha}(\omega) = f_{\alpha}(\omega+n\Omega) \delta_{nn'}\delta_{\sigma \sigma'}$ with the fermionic reservoir distribution function $f_{\alpha}(\omega) = (e^{(\omega-\mu_{\alpha})/T^{(\alpha)}_{\mathrm{res}}} + 1)^{-1}$ at temperature $T^{(\alpha)}_{\mathrm{res}}$ and chemical potential $\mu_{\alpha}$.
Accordingly, the retarded Green's function of the noninteracting system dressed with reservoirs that couple to each orbital independently is given by
\begin{align}
    (\hat G^{\mathrm{R}})^{-1}(\omega) &= (\omega + n\Omega + i\frac{\gamma}{2})\delta_{nn'}\delta_{\sigma\sigma'} - \hat H^0_{\sigma\sigma';n-n'}\;,\label{eq: Floquet Green's function}
\end{align}
where $\gamma=\sum_\alpha \gamma_\alpha$ gives the total hybridization of the reservoirs with self-energies $\Sigma^{\mathrm{R/A/K}}_{\mathrm{res}} = \sum_{\alpha} \Sigma^{\mathrm{R/A/K}}_{\alpha}$. The Fourier components of the noninteracting Hamiltonian are given by
\begin{equation}
    \hat H^0_{\sigma\sigma';n-n'} = \frac{1}{T}\int_0^T \d t \; \hat H^0_{\sigma\sigma'}(t)e^{i(n-n')\Omega t}\;.
\end{equation}
 One can also study driven reservoirs within this formalism. For example, a time dependency of the chemical potentials $\mu_{\alpha} \to\mu_{\alpha}(t)$ can be realized through a gauge transformation \cite{Eissing_2016, Honeychurch_2024}. However, we will not consider this case here.
\subsection{Functional Renormalization Group} \label{section: FRG}
The functional renormalization group~\cite{Metzner_2012, Kopiez_2012} has been used in the past for various different systems, ranging from quantum dots \cite{Jakobs_2010a,Jakobs_2010b, Ge2024} to 2D materials \cite{Honerkamp_2004, Hille_2020} in different approximations. Here we concentrate on the quantitative description of fermionic quantum dots out of equilibrium that has been established using the Keldysh formalism in Refs.~\cite{Jakobs_2010a,Jakobs_2010b, Kennes_2012, Ge2024}. Within the FRG one exactly rewrites the quantum many-body problem in terms of an infinite hierarchy of differential equations for the interacting $n$-particle vertex functions $\Gamma_n$, following the Wilsonian renormalization group idea~\cite{Wilson_1974}. The vertex functions depend on the bare propagator $g$ and are obtained as functional derivatives of the effective action, which is defined as the Legendre transform of the generating functional constructed from the Keldysh action~\cite{Wetterich_1993, Salmhofer_2001}. The differential equations are obtained by introducing a cutoff scale $\Lambda$ into the bare propagator $g\to g^{\Lambda}$ and subsequent differentiation with respect to $\Lambda$ yields the infinite coupled set of flow equations for $\Gamma_n$ in the spirit of Wilson's renormalization group
\begin{equation}
 \partial_{\Lambda} \Gamma_n^{\Lambda} = \mathcal{F}(\Gamma^{\Lambda}_1, \Gamma^{\Lambda}_2, \dots,\Gamma^{\Lambda}_{n+1})\;.   
\end{equation}
The cutoff scale is chosen such that at $\Lambda_{\mathrm{ini}}$ the system is analytically solvable (or in some reasonable approximation) and one integrates over $\Lambda$ to obtain the interacting solution of interest at $\Lambda_{\mathrm{fin}}$, regularizing possible infrared divergences throughout the flow. Since the set of equations is infinite, no exact solution is possible in practice, and one needs to resort to truncation schemes. In a common approximation, one sets all $\Gamma^{\Lambda}_{n>2}=0$ and only considers the flow of the one-particle and two-particle vertex function, which can be attributed to the self-energy $\Gamma^{\Lambda}_1 = -\Sigma^{\Lambda}$ and the effective two-particle correlation function $\Gamma^{\Lambda}_2 \equiv \Gamma^{\Lambda}$ (with amputated legs in respective diagram). Introducing the so-called single-scale propagator 
\begin{equation}
    S_{\Lambda} = G_{\Lambda}g_{\Lambda}^{-1}\frac{\d g_{\Lambda}}{\d \Lambda}g_{\Lambda}^{-1}G_{\Lambda}\;,
\end{equation} 
and the differentiated bubble diagram
\begin{equation}
   \partial_{\Lambda} \Pi^{34|3'4'}_{\Lambda} = \partial_{\Lambda} [G^{3|3'}_{\Lambda}G^{4|4'}_{\Lambda}]=G^{3|3'}_{\Lambda}S^{4|4'}_{\Lambda} + S^{3|3'}_{\Lambda} G^{4|4'}_{\Lambda}\;, \label{eq:diff bubble}
\end{equation} 
the flow equations up to $n=2$ read
\begin{align}
    \partial_{\Lambda}\Sigma^{1'|1}_{\Lambda} &= -\frac{i}{2\pi}\Gamma^{1'2'|12}_{\Lambda}S_{\Lambda}^{2|2'}\;, \label{eq:selfen_ode}\\
    \partial_{\Lambda}\Gamma^{1'2'|12}_{\Lambda} &= \frac{i}{4\pi}\Gamma^{1'2'|34}_{\Lambda}\partial_{\Lambda}\Pi_{\Lambda}^{34|3'4'}\Gamma^{3'4'|12}_{\Lambda} \nonumber\\
    &+ \frac{i}{2\pi}\Gamma^{1'4'|32}_{\Lambda}\partial_{\Lambda}\Pi_{\Lambda}^{34|3'4'}\Gamma^{3'2'|14}_{\Lambda} \nonumber\\
    &- \frac{i}{2\pi}\Gamma^{1'3'|14}_{\Lambda}\partial_{\Lambda}\Pi_{\Lambda}^{34|3'4'}\Gamma^{4'2'|32}_{\Lambda} \;. \label{eq:vertex_ode}
\end{align}
For the differentiated bubble, we also employ the Katanin substitution~\cite{Katanin_2004} and replace
\begin{equation}
    S_{\Lambda}\to\partial_{\Lambda}G = S_{\Lambda} + G_{\Lambda}[\partial_{\Lambda}\Sigma_{\Lambda}] G_{\Lambda}\;,
\end{equation}
 in Eq.~\eqref{eq:diff bubble}.
The three terms in Eq.~\eqref{eq:vertex_ode} can be attributed to the particle-particle ($P$), particle-hole ($C$) and direct particle-hole ($D$) channel known from the ladder diagrams one computes in the Bethe-Salpeter equations within random-phase approximation (RPA) \cite{Salpeter_1951, Martin_2016}. Thus, a one-loop FRG setup like this has an inherent bias towards ladder-type diagrams and the associated divergences. In extended systems, one often employs a static approximation, neglecting all frequency dependencies of the two-particle vertex and sometimes also the full self-energy flow. Here, we want to keep the frequency dependency at least partially to investigate the impact of inelastic scattering on Floquet engineering. 
The frequency-dependent Floquet FRG equations are obtained from the real-time version in Eqs.~\eqref{eq:selfen_ode} and \eqref{eq:vertex_ode} (dropping the explicit $\Lambda$-dependency) as 
\begin{widetext}
\begin{align}
    \partial_{\Lambda}\Sigma^{1'|1}_{n}(\omega) = &-\frac{i}{2\pi}\sum_{l}\int_{-\infty}^{\infty}\d \tilde\omega \;\Gamma^{1'2'|12}_{l}\left(\omega + \tilde\omega, \omega-\tilde\omega, (\frac{n}{2} - \frac{l}{4})\Omega\right)S_{n-l}^{2|2'}(\tilde\omega)\;\label{eq:floquet_selfen_ode}, \\
     \partial_{\Lambda} P_n^{1'2'|12}(\omega, \nu, \nu') = &\frac{i}{4\pi}\sum_{k_1, k_2, k_3}\int \d\tilde\nu\;\Gamma^{1'2'|34}_{n - k_1 - k_2 - k_3}(\omega + \frac{\Omega}{2}(k_1 + k_2 + k_3), \nu - \frac{\Omega}{2}(k_1 - k_2), \tilde\nu) \nonumber \\
     &\times \partial_{\Lambda}\Pi^{34|3'4'}_{k_1,k_2}(\frac{\omega}{2} - \tilde\nu - \frac{\Omega}{2}(\frac{n}{2} - k_1 - k_3),\frac{\omega}{2} + \tilde\nu - \frac{\Omega}{2}(\frac{n}{2} - k_2 - k_3)) \nonumber \\
     &\times \Gamma^{3'4'|12}_{k_3}(\omega - \frac{\Omega}{2}(n - k_3), \tilde\nu,\nu')\;,
     \label{eq: Frequency Dependent Floquet P channel}\\
     \partial_{\Lambda} C_n^{1'2'|12}(\omega, \nu, \nu') = &\frac{i}{2\pi}\sum_{k_1, k_2, k_3}\int \d\tilde\nu\;\Gamma^{1'4'|32}_{n - k_1 - k_2 - k_3}(\omega + \frac{\Omega}{2}(k_1 + k_2 + k_3), \nu - \frac{\Omega}{2}(k_1 - k_2), \tilde\nu) \nonumber \\
     &\times \partial_{\Lambda}\Pi^{34|3'4'}_{k_1,k_2}( \tilde\nu -\frac{\omega}{2} - \frac{\Omega}{2}(\frac{n}{2} - k_1 - k_3),\frac{\omega}{2} + \tilde\nu + \frac{\Omega}{2}(\frac{n}{2} - k_2 - k_3)) \nonumber \\
     &\times \Gamma^{3'2'|14}_{k_3}(\omega - \frac{\Omega}{2}(n - k_3), \tilde\nu,\nu')\;, 
     \label{eq: Frequency Dependent Floquet C channel}\\
     \partial_{\Lambda} D_n^{1'2'|12}(\omega, \nu, \nu') = &-\frac{i}{2\pi}\sum_{k_1, k_2, k_3}\int \d\tilde\nu\;\Gamma^{1'3'|14}_{n - k_1 - k_2 - k_3}(\omega + \frac{\Omega}{2}(k_1 + k_2 + k_3), \nu - \frac{\Omega}{2}(k_1 - k_2), \tilde\nu) \nonumber \\
     &\times \partial_{\Lambda}\Pi^{34|3'4'}_{k_1,k_2}( \tilde\nu -\frac{\omega}{2} - \frac{\Omega}{2}(\frac{n}{2} - k_1 - k_3),\frac{\omega}{2} + \tilde\nu + \frac{\Omega}{2}(\frac{n}{2} - k_2 - k_3)) \nonumber \\
     &\times \Gamma^{4'2'|32}_{k_3}(\omega - \frac{\Omega}{2}(n - k_3), \tilde\nu,\nu') \;.
     \label{eq: Frequency Dependent Floquet D channel}
\end{align}
\end{widetext}
The vertices are written in their channel-native frequency parametrization from Ref.~\cite{Ge2024} and the superscript indices now only contain the Keldysh and orbital indices. The differentiated bubble is written in terms of propagators in single index form, which is beneficial for the numerics but can of course be rewritten to Floquet matrix form, which is of advantage when the frequency integration can be carried out analytically (this is possible if one only considers static self-energy feedback or no self-energy feedback at all, as well as the approximations of the vertices we discuss next). In this work, we will employ further approximations to the frequency dependence of the vertex, as the combination with the Floquet degree of freedom makes numerical computations exhaustively expensive. The number of integrals to evaluate scales with $n_f^4\times n^3_{\omega}$, where $n_{\omega}\sim 10^2$ the number of frequency points required for good resolution and $n_f\sim 10^1$ the number of Floquet coefficients required for convergence in the Floquet indices.
\subsubsection{Channel decomposition}
Following Ref.~\cite{Jakobs_2010b}, we approximate the two-particle vertex as a frequency decomposition into the three ``channel-native'' components and only keep the dependency on the respective bosonic frequencies
\begin{align}
    \Gamma_{\Lambda}^{1'2'|12} (\omega_P,\omega_C,\omega_D) &\approx \Gamma_{0}^{1'2'|12} + P_{\Lambda}^{1'2'|12} (\omega_P) \nonumber\\ 
    &+ C_{\Lambda}^{1'2'|12} (\omega_C) +D_{\Lambda}^{1'2'|12} (\omega_D)\;,
\end{align}
where $\Gamma_{0}=\Gamma_{\Lambda \to \infty}$ is the initial value of the vertex (usually the bare interaction or an easily obtainable approximation). The vertex flow equations still mix the different channels according to the heuristic approach introduced in Ref.~\cite{Jakobs_2010a}, where static interaction vertices $\Phi$ are attributed to the contributions of the other channels.
\begin{align}
    P_{\Lambda}^{1'2'|12} (\omega)\to &P_{\Lambda}^{1'2'|12} (\omega) + (\Phi^{\mathrm C}_{\Lambda})^{1'2'|12} + (\Phi^{\mathrm D}_{\Lambda})^{1'2'|12} \;, \\
    C_{\Lambda}^{1'2'|12} (\omega)\to &C_{\Lambda}^{1'2'|12} (\omega) + (\Phi^{\mathrm P}_{\Lambda})^{1'2'|12} + (\Phi^{\mathrm D}_{\Lambda})^{1'2'|12} \;, \\
    D_{\Lambda}^{1'2'|12} (\omega)\to &D_{\Lambda}^{1'2'|12} (\omega) + (\Phi^{\mathrm P}_{\Lambda})^{1'2'|12} + (\Phi^{\mathrm C}_{\Lambda})^{1'2'|12} \;.
\end{align}
We found that this approach does not work well for the driven systems as it does not consistently treat the mixing of the Floquet components and therefore set the static mixing terms to zero, which decouples the channels and leads to an RPA-like resummation.
In the equilibrium SIAM this would lead to a divergence in the channels which is circumvented by keeping the bath temperature finite and employing the Katanin substitution that partially includes multiloop diagrams (see the discussion in Appendix~\ref{app: equilibrium benchmark}). Keeping the self-energy feedback, it can be considered an RG analogue of the self-consistent fluctuation-exchange (FLEX) approximation  \cite{White_1992, Schl_nzen_2019, Honeychurch_2024}. Note, however, that the FRG is only conserving up to $\mathcal{O}(U^2)$ [$\mathcal{O}(U^4)$ with Katanin substitution], in contrast to FLEX, which is $\phi$-derivable~\cite{Baym_1961,Bickers_1989, Bickers_1991}. 
Overall, the approximations restrict the calculations to an accuracy of $\mathcal{O}(U^2)$ instead of $\mathcal{O}(U^3)$ that one can achieve with the full frequency parametrization. Therefore, quantitatively accurate results can only be expected for small to intermediate interactions.\\
The flow equations for the channels now simplify significantly as each vertex is a single frequency object like the Green's function, so we can write them as a Floquet matrix (now the hat indicates only the Floquet indices and we dropped the orbital indices for readability). We can also attribute retarded, advanced, and Keldysh components in the vertex Keldysh structure \cite{Jakobs_2010} as
\begin{align}
   \hat P^{\mathrm R}(\omega) =  \hat P^{12|22}(\omega)\;,\;\;\;\;\hat P^{\mathrm K}(\omega) &=  \hat P^{12|12}(\omega)\;,\\
   \hat C^{\mathrm R}(\omega) =  \hat C^{12|22}(\omega)\;,\;\;\;\;\hat C^{\mathrm K}(\omega) &=  \hat C^{12|12}(\omega)\;,\\
   \hat D^{\mathrm R}(\omega) =  \hat D^{12|22}(\omega)\;,\;\;\;\;\hat D^{\mathrm K}(\omega) &=  \hat D^{12|21}(\omega)\;.
\end{align}
Within this approximation the flow equations can be restructured in a somewhat simpler form in terms of (differentiated) particle-hole and particle-particle susceptibilities
\begin{widetext}
    
\begin{align}
    (\partial_{\Lambda}\chi^{\sigma_1\sigma_2|\sigma_1'\sigma_2'}_{\mathrm{ph}})^{\mathrm R}_n(\omega) &= i\sum_m\int \frac{\d \omega'}{2\pi } \partial_{\Lambda}\left[ \Pi^{21|22} + \Pi^{22|12}\right]^{\sigma_1\sigma_2|\sigma_1'\sigma_2'}_{m,n-m}(\omega' - \frac{\omega}{2},\omega' + \frac{\omega}{2})\;, \label{eq:ph_susceptibility}\\
    (\partial_{\Lambda}\chi^{\sigma_1\sigma_2|\sigma_1'\sigma_2'}_{\mathrm{ph}})^{\mathrm K}_n(\omega) &= i\sum_m\int \frac{\d \omega'}{2\pi }\partial_{\Lambda}\left[ \Pi^{12|21} + \Pi^{21|12} + \Pi^{22|22}\right]^{\sigma_1\sigma_2|\sigma_1'\sigma_2'}_{m,n-m}(\omega' - \frac{\omega}{2},\omega' + \frac{\omega}{2})\;, \\
    (\partial_{\Lambda}\chi^{\sigma_1\sigma_2|\sigma_1'\sigma_2'}_{\mathrm{pp}})^{\mathrm R}_n(\omega) &= i\sum_m\int \frac{\d \omega'}{2\pi } \partial_{\Lambda}\left[ \Pi^{22|12} + \Pi^{22|21}\right]^{\sigma_1\sigma_2|\sigma_1'\sigma_2'}_{m,n-m}(\frac{\omega}{2} +\omega',\frac{\omega}{2}- \omega' )\;, \\
    (\partial_{\Lambda}\chi^{\sigma_1\sigma_2|\sigma_1'\sigma_2'}_{\mathrm{pp}})^{\mathrm K}_n(\omega) &= i\sum_m\int \frac{\d \omega'}{2\pi }\partial_{\Lambda}\left[ \Pi^{22|11} + \Pi^{11|22} + \Pi^{22|22}\right]^{\sigma_1\sigma_2|\sigma_1'\sigma_2'}_{m,n-m}(\frac{\omega}{2} +\omega',\frac{\omega}{2}- \omega')\;,\label{eq:pp_susceptibility}
\end{align}
\end{widetext}
where we have explicitly executed the Keldysh summation. The susceptibilities as well as the vertices can, in this approximation, be transformed into Floquet matrices, turning the actual computation of the right-hand side of the vertex flow equations into batched matrix products. Leaving any other degrees of freedom implicit [which still follow the summation rule in Eq.\eqref{eq:vertex_ode}] we have
    \begin{align}
         \partial_{\Lambda}\hat X^{\mathrm R}(\omega) = &\;\xi_{X}(\hat \Gamma_0 + \hat X^{\mathrm R}(\omega) )\left[\partial_{\Lambda}\hat\chi_{\xi}^{\mathrm R}\right](\omega)(\hat \Gamma_0 + \hat X^{\mathrm R}(\omega))\;,  \label{eq:RPA_ode_R}\\
         \partial_{\Lambda}\hat X^{\mathrm K}(\omega) = &\;\xi_X(\hat \Gamma_0 + \hat X^{\mathrm R}(\omega) )\left[\partial_{\Lambda}\hat\chi_{\xi}^{\mathrm K}\right](\omega)(\hat \Gamma^{\dagger}_0 + \hat X^{\mathrm A}(\omega)) \nonumber\\ 
         &+ \xi_X(\hat \Gamma_0 + \hat X^{\mathrm R}(\omega) )\left[\partial_{\Lambda}\hat\chi_{\xi}^{\mathrm R}\right](\omega) \hat X^{\mathrm K}(\omega) \nonumber\\
         &+ \xi_X\hat X^{\mathrm K}(\omega) \left[\partial_{\Lambda}\hat\chi_{\xi}^{\mathrm A}\right](\omega)(\hat \Gamma^{\dagger}_0 + \hat X^{\mathrm A}(\omega))\;,\label{eq:RPA_ode_K}
    \end{align}
 for $X\in\{P,C\}$,
where we introduced shorthands for the channel specific prefactors $\xi_P=1/2$, $\xi_C=1$, and $\hat\Gamma_0\equiv \hat\Gamma_0^{12|22}$ is the only independent Keldysh component of the bare vertex. 
The Keldysh summation of the $D$-channel differs slightly and is given by
\begin{align}
         \partial_{\Lambda}\hat D^{\mathrm R}(\omega) = &-(\hat \Gamma_0 + \hat D^{\mathrm R}(\omega) )\left[\partial_{\Lambda}\hat\chi_{\mathrm{ph}}^{\mathrm R}\right](\omega)(\hat \Gamma_0 + \hat D^{\mathrm R}(\omega))\;, \label{eq:D_RPA_ode_R}\\
             \partial_{\Lambda}\hat D^{\mathrm K}(\omega) = &-(\hat \Gamma_0 + \hat D^{\mathrm R}(\omega) )\left[\partial_{\Lambda}\hat\chi_{\mathrm{ph}}^{\mathrm K}\right](\omega)(\hat \Gamma^{\dagger}_0 + \hat D^{\mathrm A}(\omega)) \nonumber\\ 
         &-(\hat \Gamma_0 + \hat D^{\mathrm R}(\omega) )\left[\partial_{\Lambda}\hat\chi_{\mathrm{ph}}^{\mathrm A}\right](\omega)\hat D^{\mathrm K}(\omega) \nonumber\\
         &-\hat D^{\mathrm K}(\omega) \left[\partial_{\Lambda}\hat\chi_{\mathrm{ph}}^{\mathrm R}\right](\omega)(\hat \Gamma^{\dagger}_0 + \hat D^{\mathrm A}(\omega))\;. \label{eq:D_RPA_ode_K}
\end{align}

Together with Eq.~\eqref{eq:floquet_selfen_ode}, Eqs.~\eqref{eq:RPA_ode_R}-\eqref{eq:D_RPA_ode_K} give the set of coupled flow equations that we solve numerically [the Keldysh sum in Eq.~\eqref{eq:floquet_selfen_ode} has been carried out in Ref.~\cite{Jakobs_2010a}].

\begin{figure*}[ht!]
\centering
\begin{tikzpicture}[baseline=(table.north)]

    \node (table) {
        \begin{tabular}{c|c}
        Flow scheme  & Flowing quantities \\ \hline\hline
        FRG & $\Sigma_n(\omega)$, $P_n(\omega)$, $C_n(\omega)$,  $D_n(\omega)$, on frequency grid defined by Eq.~\eqref{eq: root grid} \\ 
        rsFRG &  $\Sigma_n(\omega)$, $P_n(\omega)$, $C_n(\omega)$,  $D_n(\omega)$, for $\omega\in\{-n_f^{(2)}\Omega,-(n_f^{(2)}-1/2)\Omega,\dots,(n_f^{(2)}-1/2)\Omega, n_f^{(2)}\Omega\}$  \\ 
        sFRG &   $\Sigma_n$, $P_n$, $C_n$,  $D_n$, constant for all $\omega$ \\ 
        \end{tabular}
    };

    \node[left=.1cm of table, yshift=-1.1cm, rotate=270] (arrowL) {
        \begin{tikzpicture}
            \node[
                single arrow,
                single arrow head extend=3pt,
                minimum width=.1cm,      
                minimum height=1.5cm,     
                draw,
                fill=blue!25
            ] {};
        \end{tikzpicture}
    };
    \node[left=1cm of arrowL, yshift=.3cm, rotate=90] {\textbf{Performance}};

    \node[right=.1cm of table, yshift=-1.0cm, rotate=90] (arrowR) {
        \begin{tikzpicture}
            \node[
                single arrow,
                single arrow head extend=3pt,
                minimum width=.1cm,      
                minimum height=1.5cm,     
                draw,
                fill=orange!55
            ] {};
        \end{tikzpicture}
    };
    \node[right=1cm of arrowR, yshift=-1.8cm, rotate=90] {\textbf{Accuracy}};

\end{tikzpicture}

\caption{The different approximations to the FRG equations investigated in this work. The most elaborate scheme (FRG) truncates the flow equations at second order and separates the channels (no inter-channel mixing). The second scheme (rsFRG) reduces the numerical effort by an order of magnitude by only evaluating quantities on a discrete frequency grid of multiples of $\Omega/2$, the replica grid. In between, the quantities are linearly interpolated. Here, $n_f^{(2)}$ defines the maximal Floquet coefficient kept in all flowing quantities. The last and least computationally intensive scheme approximates all quantities as fully static in $\omega$ (sFRG).}
\label{fig:FRG approximation table}
\end{figure*}
\subsubsection{Additional approximations}
As we develop this framework with the goal in mind to eventually simulate extended systems and real quantum materials under Floquet driving, we also test additional approximations. Benchmarking these cheaper approximations will give a good estimate of whether one can expect good results using them on extended systems, as well. A so-called static approximation, in which we neglect the dependency on $\omega$ in all quantities, is regularly done in equilibrium simulations~\cite{Profe_2022,Profe_2024} at the expense of quantitative correctness. Still, free energy and Fermi surface arguments allow for the qualitative implications of such static simulations to remain accurate. In non-equilibrium, these arguments do not generally hold. Therefore, we test numerically whether for the Floquet steady-state we can still get qualitatively accurate results for static quantities like the effective interaction
\begin{equation}
    V_n\equiv\Gamma^{\mathrm R}_n(0,0,0)\;. \label{eq: effective interaction}
\end{equation}
We test two different possibilities, i) the fully static approximation, where we  only keep the Floquet coefficients, denoted ``sFRG'', and ii) a static approximation that takes into account frequency shifts by multiples of $\Omega/2$ by setting up a discretized replica-frequency grid at exactly these values. This is motivated by considering Eqs.~\eqref{eq:di_to_si},\eqref{eq:si_to_di} which require these multiples to work correctly for static quantities at $\omega=0$. Physically, this can be interpreted as including all Floquet replica effects at $\omega=0$. We denote ii) by ``rsFRG''. A summary of the approximations is shown in Fig.~\ref{fig:FRG approximation table}.

\subsubsection{Hybridization flow}
The truncation of the flow equations causes the choice of the regulator, which should in principle be arbitrary, to become relevant. The extensive discussion in Ref.~\cite{Jakobs_2010a} showed that causality and grand-canonical ensemble statistics might be altered in unphysical ways if one considers, for example, sharp cutoff schemes, as it is frequently done in equilibrium \cite{Andergassen_2004, Meden_2008, Karrasch_2008, Hille_2020, Profe_2024}. Thus, we opt for the hybridization flow presented in Ref.~\cite{Jakobs_2010a}, called $\Delta$-flow in Ref.~\cite{Ge2024}. The flow can be interpreted as a successive decoupling from an auxiliary reservoir, which at the beginning $\Lambda\to\infty$, is coupled to the system with infinite hybridization strength. We thus introduce the flow parameter as 
\begin{equation}
    \gamma/2 \to (\gamma + \Lambda)/2 \equiv \Delta \;.
\end{equation}
The single-scale propagator is then given by
\begin{align}
    \hat S^{\mathrm R}(\omega) =& -\frac{i}{2} \left[\hat G^{\mathrm R}\right]^2(\omega)\;,\\
    \hat S^{\mathrm K}(\omega) =& -\frac{i}{2} \hat G^{\mathrm R}\hat G^{\mathrm K} + \frac{i}{2} \hat G^{\mathrm K} \hat G^{\mathrm A} \nonumber\\
    &-i\hat G^{\mathrm R}\left[\mathbb{1} - 2\hat f(\omega)\right] \hat G^{\mathrm A} \;,
\end{align}
The initial conditions for the flow are given by the Hartree self-energy and the bare interaction
\begin{align}
    \hat \Sigma^{\mathrm R}_{\Lambda\to\infty} &= \hat\Sigma^{\mathrm H}\;,\\
    \hat\Gamma^{1'2'|12}_{\Lambda\to\infty} &= \hat\Gamma^{1'2'|12}_0\;,
\end{align}
all other terms are initialized to zero. To obtain the Hartree self-energy, we perform a self-consistent Hartree-Fock computation (see Section~\ref{sec:SCHF}).
The hybridization flow, as introduced here, has the benefit of describing a physical system at each point during the flow and, furthermore, being equivalent to an interaction flow in $U/\Delta$ (in the case of undriven reservoirs, and the temperature of both the physical reservoir and the auxiliary being set equal).
This enables a further reduction in the numerical cost of the flow by initializing not with the Hartree value, but with the RPA solution at small $U/\Delta\sim0.1$ (see Section~\ref{sec:RPA}). For smaller $U/\Delta$, the RPA and FRG results should give the same results. This can reduce the initial value $\Lambda_{\mathrm{ini}}$ needed for converged results by several orders of magnitude.
For Floquet Green's functions specifically, the RPA initialization is practically required for the hybridization flow to work, since the number of Floquet components that need to be retained for convergence scales with $\Delta/\Omega$ (see Section~\ref{sec: representation of Floquet GF}). This makes the Floquet Green's functions hard to resolve at large initial scales. It might be of relevance to investigate tailored Floquet cutoffs that avoid this pitfall. One way would be to set $\Omega\to \Delta\Omega$ so that $\Omega/\Delta$ stays fixed during the flow. However, this further complicates the single-scale propagator, as one has to explicitly differentiate $\Omega$ which occurs in $\hat{G}_{\Lambda}$. This will be a matter of future work. For extended systems, one could also use a momentum cutoff scheme \cite{Metzner_2012}.

\subsubsection{Spin summation for the $\mathrm{SU}(2)$-symmetric SIAM}
In the case of the SIAM with no external magnetic field and metallic leads, we can also solve the summation over spin indices explicitly to further simplify the flow equations. Then, the Green's functions reduce to a single number per Keldysh component, frequency point, and Floquet index. For the vertices, we keep the $\hat X_{\uparrow\downarrow|\uparrow\downarrow}\equiv\hat X_{\uparrow\downarrow}$ components in each channel and the other components are related through $\hat \Gamma_{\uparrow\uparrow} = \hat \Gamma_{\uparrow\downarrow} + \hat \Gamma_{\downarrow\uparrow}$. 
For the $P$ and $C$ channels, we can drop the spin summation entirely (the $P$ channel gets a trivial factor of $2$ from the spin summation), while for the $D$ channel and the self-energy one can perform the sum explicitly using crossing symmetry $\hat D_{\downarrow\uparrow}(\omega) = - \hat{C}_{\uparrow\downarrow}(-\omega)$ (see Ref.~\cite{Jakobs_2010a}).

\subsection{Alternative approaches} \label{sec:alternative_approaches}
There are several alternatives that one can consider that are on the same level of perturbative approximation in $U/\Delta$. Here we benchmark the FRG against bare second order perturbation theory (2PT) and the self-consistent GW approximation that attempts to incorporate screening effects \cite{Schl_nzen_2019}. Both of these approaches only contain the particle-hole susceptibility, and thus only contain a subset of the diagrams that are resolved by the FRG. Still, at small $U/\Delta$ we expect a good agreement between all these approaches. 

\subsubsection{Self-consistent Hartree-Fock} \label{sec:SCHF}
Both the 2PT and the FRG require a self-consistent Hartree-Fock (SCHF) computation as the initial value for the self-energy.
The SCHF approximation aims at a self-consistent solution of only the Hartree and Fock diagrams, which leads to a frequency-independent self-energy. In Ref.~\cite{Eissing_2016} a Floquet FRG scheme was introduced that only flows the self-energy, which can be considered an RG enhancement to SCHF. For the $\mathrm{SU}(2)$-symmetric SIAM, the SCHF consists of computing the Hartree self-energy 
\begin{equation}
    \Sigma^{\mathrm H}_{n} = \frac{U}{2\pi i} \int_{-\infty}^{\infty} \d \omega \;G^{<}_n(\omega)\;,
\end{equation}
where $G^{<} = \frac{1}{2}(G^{\mathrm A} - G^{\mathrm R} + G^{\mathrm K})$ is the lesser Green's function, iteratively reinserting the self-energy into the right-hand side until convergence is reached~\cite{Ge2024}. 
\subsubsection{Bare second order perturbation theory}
We implement the bare 2PT by starting with a SCHF computation and subsequently dress the Green's functions with $\Sigma^{\mathrm{H}}$ to compute the non-differentiated version of Eq.~\eqref{eq:ph_susceptibility} to obtain the particle-hole susceptibility. The vertex in 2PT is given by Eq.~\eqref{eq:RPA_ode_R} (plugging in the full susceptibility) and setting $\hat X=0$, i.e., only contracting with the bare vertex $\hat \Gamma_0$. 
\begin{align}
    \hat V_{\mathrm{2PT}}^{\mathrm R}(\omega) = & \hat \Gamma_0\hat\chi^{\mathrm R}_{\mathrm{ph}}\hat \Gamma_0\;, \\
    \hat V_{\mathrm{2PT}}^{\mathrm K}(\omega) = & \hat \Gamma_0\hat\chi^{\mathrm K}_{\mathrm{ph}}\hat \Gamma_0\;.
\end{align}
This vertex can then be plugged into the non-differentiated version of Eq.~\eqref{eq:selfen_ode} (replacing the single-scale propagator by the full propagator $S\to G$) to obtain the self-energy within 2PT that is then given by $\Sigma^{\mathrm{corr}} = \Sigma^{\mathrm{H}} + \Sigma^{\mathrm{2PT}}$. Note that this approach does not enforce any self-consistency on the two-particle level.

\subsubsection{Random-phase approximation} \label{sec:RPA}
It is possible to relate the FRG flow equations to the RPA by keeping the self-energy fixed at the Hartree level and only flowing Eqs.~\eqref{eq:RPA_ode_R}-\eqref{eq:D_RPA_ode_K} without channel-mixing. One can then solve the integral over $\Lambda$ analytically to obtain the geometric series known from the ladder approximation. In the particle-hole channel, this yields a solution in terms of a Floquet matrix inverse 
\begin{align}
    \hat V_{\mathrm{RPA}}^{\mathrm R}(\omega) = &\left(\mathbb{1} - \hat \Gamma_0 \hat\chi^{\mathrm R}_{\mathrm{ph}}\right)^{-1}  \hat V^{\mathrm R}_{\mathrm{2PT}}\;, \\
    \hat V_{\mathrm{RPA}}^{\mathrm K}(\omega) = &\left(\mathbb{1} - \hat \Gamma_0\hat\chi^{\mathrm R}_{\mathrm{ph}}\right)^{-1}  \nonumber \\
    &\times\left(\hat V^{\mathrm K}_{\mathrm{2PT}} + \hat \Gamma_0 \hat\chi^{\mathrm K}_{\mathrm{ph}}\hat V^{\mathrm A}_{\mathrm{2PT}}\right)\;.
\end{align} 
This equation also holds for multiple orbitals by carefully batching the four orbital indices of the generalized susceptibilities to obtain the corresponding matrices in a combined orbital-Floquet space. The RPA equations have already been employed to study Floquet-engineered phase-diagrams in 2D systems~\cite{Herre_2026}. This approach does not regulate divergences and in equilibrium, as well as in the driven settings discussed here, it breaks down for $U/\Delta\gtrsim\pi$~\cite{Jakobs_2010} and we do not present bare RPA results in this work. However, they provide a better starting point for the FRG than only SCHF, and we use the RPA vertex, as well as the resulting self-energy $\Sigma^{\mathrm{corr}} = \Sigma^{\mathrm{H}} + \Sigma^{\mathrm{RPA}}$ as the initial value for the FRG to decrease $\Lambda_{\mathrm{ini}}$ and reduce numerical effort. 

\subsubsection{GW approximation}
Within the GW approximation, we enforce both self-consistency of the full frequency-dependent self-energy and incorporate screening effects by writing diagrams including the 2PT vertex instead of the bare vertex. Although the GW approximation as implemented here is derived from Hedin's equations \cite{Hedin_1965, Schl_nzen_2019} as a conserving approximation from a functional $\Phi[G]$, the resulting equation for the vertex is still a ladder diagram similar to the particle-hole channel in RPA, replacing the bare vertex with the 2PT vertex.
\begin{align}
    \hat V_{\mathrm{GW}}^{\mathrm R}(\omega) = &\left(\mathbb{1} - \hat V^{\mathrm R}_{\mathrm{2PT}} \hat\chi^{\mathrm R}_{\mathrm{ph}}\right)^{-1}  \hat V^{\mathrm R}_{\mathrm{2PT}}\;, \\
    \hat V_{\mathrm{GW}}^{\mathrm K}(\omega) = &\left(\mathbb{1} - \hat V^{\mathrm R}_{\mathrm{2PT}} \hat\chi^{\mathrm R}_{\mathrm{ph}}\right)^{-1}  \nonumber \\
    &\times\left(\hat V^{\mathrm K}_{\mathrm{2PT}} + \hat V^{\mathrm R}_{\mathrm{2PT}}\hat\chi^{\mathrm K}_{\mathrm{ph}}\hat V^{\mathrm A}_{\mathrm{2PT}} + \hat V^{\mathrm K}_{\mathrm{2PT}}\hat\chi^{\mathrm A}_{\mathrm{ph}}\hat V^{\mathrm A}_{\mathrm{2PT}}\right)\;.
\end{align} 
Again, this works for a general multi-orbital system as well by batching the four orbital indices correctly to transform all objects into matrices (see Ref.~\cite{Schl_nzen_2019} for the explicit contractions). 
In contrast to the 2PT computation, where we solve for the self-energy once with SCHF-dressed Green's functions, in the GW the Hartree contribution is solved for self-consistency alongside the GW part $\Sigma^{\mathrm{corr}} = \Sigma^{\mathrm{H}} + \Sigma^{\mathrm{GW}}$.
In equilibrium, the GW approximation has been applied to the SIAM in Ref.~\cite{Wang_2008}. One could also employ the FLEX approximation as presented in Ref.~\cite{Honeychurch_2024}, however, it suffers from the same divergences as the RPA, since it includes the RPA particle-hole channel. Thus, it requires great care in the self-consistency loop to reach $U/\Delta \gtrsim \pi$ with questionable results for the SIAM in equilibrium~\cite{White_1992} and we do not present it here.

\subsection{Numerical implementation}\label{section: Numerical implementation}
The implementation of the Floquet-Keldysh diagrammatics requires an efficient high-performance implementation, with the most expensive objects to compute being the susceptibilities in Eqs.~\eqref{eq:ph_susceptibility}-\eqref{eq:pp_susceptibility}. Further, the implementation of the Floquet Green's functions themselves requires some care to consistently switch between the single and double index representation as needed. On additional technical grounds, the code requires i) an efficiently parallelized adaptive quadrature routine to solve in parallel for $\mathcal{O}(n_{\sigma}^4n_f^2n_{\omega}^2)$ integrals, ii) a self-consistency solver with an adequate mixing scheme (we use simple Anderson mixing which leaves room for further improvement), iii) an adaptive ODE solver that can reliably solve the in principle non-stiff FRG equations. Here, we use a Runge-Kutta 4(5) stepper. The code is implemented using the \textit{jax} python framework, which we found to provide the best compromise between performance and development speed with efficient GPU acceleration. This is necessary to simulate more than few-level quantum dot systems, the runtimes we could achieve being $\sim 5-10$h on a single NVIDIA H100 GPU for fully frequency-dependent runs.

\subsubsection{Representation of Floquet Green's functions}\label{sec: representation of Floquet GF}
To obtain the Floquet Green's function in single index form, one needs to first generate them in double index form from Eqs.~\eqref{eq: Floquet Green's function}, which only requires knowledge of the Fourier coefficients of the time-dependent system Hamiltonian. As the Floquet matrix of the Green's function is only defined in the fundamental frequency domain $F$, the conversion to the single index version essentially consists of ``stitching'' together the matrix elements of the Floquet matrix along the real frequency axis for each Floquet coefficient, using the transformation rule in Eq.~\eqref{eq:di_to_si}. In principle, the Floquet matrix is infinitely large and covers the whole frequency axis, but, in practice, one can only keep a finite number of $n_f$ Floquet indices. Thus, for a given $n_f$, one can only construct the Floquet coefficient version of the Floquet Green's function for frequencies $|\omega|\lesssim n_f \Omega$.  This leads to numerical inaccuracies even for large $n_f$, since the Floquet Green's functions, as their equilibrium counter part, decay algebraically as $\omega^{-|\beta|}$ for $\omega\to\pm\infty$. To overcome this and also reduce the numerical effort by reducing the required $n_f$ we employ an extrapolation scheme and fit an algebraic tail of the form $a\omega^{-|\beta|}$ to each Green's function. 
In this way, we can separate $n_f$ into two separate quantities. The first is the number of Floquet coefficients required to resolve each Floquet coefficient $G_n$ on the real frequency axis, to ensure that the extrapolation works well. We set this to $n^{(1)}_f=35 \Omega /\gamma$, which suffices to resolve all relevant features in the central frequency region around $\omega=\mu$ and allow an accurate algebraic extrapolation for larger frequencies.
This works because Floquet induced features far away from $\omega=\mu$ are generally suppressed, following the envelope of the algebraic tails.
The second quantity, $n^{(2)}_f$, now only needs to be large enough to resolve the drive protocol. It scales with $n^{(2)}_f\sim A/\Omega$, where $A$ is the drive amplitude, but generally also depends on the complexity of the protocol. The convergence in $n^{(2)}_f$ must be checked for each set of parameters but can be up to a factor of $10$ smaller than $n^{(1)}_f$ for simple driving schemes.\\
Finally, to further reduce the numerical effort and memory requirements, we resolve the Floquet Green's functions on a non-uniform frequency grid as presented in Ref.~\cite{Ge2024}. For that, we define the auxiliary variable $\tilde \omega \in[-1,1]$ and map it to a root grid by 
\begin{equation}
    \omega = \mu + \frac{B\tilde\omega |\tilde\omega|}{\sqrt{1 - \tilde\omega^2}}\;, \label{eq: root grid}
\end{equation}
with a heuristically determined scale factor $B>0$.
This reduces the number of frequency points needed to resolve the features around $\omega=\mu$ and the heavy tails by another factor of $10$ to $N_{\omega}\sim10^2$. For the model and drive protocol considered in this work, it turns out that for an appropriate resolution of Floquet induced features located at $\omega=\mu\pm n\Omega$, a similar number of grid points is needed as in the equilibrium case.

\subsubsection{Efficient integration of Floquet susceptibilities}
To compute the susceptibilities in Eqs.~\eqref{eq:ph_susceptibility}-\eqref{eq:pp_susceptibility} in a memory- and compute-efficient way, we utilize the previously introduced frequency grids and extrapolation schemes to set up an adaptive quadrature routine that linearly interpolates the required $G_n$ for any frequency value on the real axis. 
This allows for guaranteed accuracy of the integration, even with the very complicated features induced by the driving. We use a Gauss-Kronrod routine with 21 nodes (GK21) that is vectorized to compute all the components of the susceptibilities in parallel. The evaluation intervals are adaptively chosen to minimize the error of the Frobenius norm of all components. Adaptive quadratures are generally not very well parallelizable beyond the function evaluation itself, and thus are not considered a prime candidate for GPU acceleration. Still, using aggressive XLA kernel fusion provided by the \textit{jax} python framework, we could implement a version compiled to NVIDIA GPUs that could get the integration time from minutes down to a few seconds. This efficiency is paramount, as the solution of the FRG differential equations typically requires $10^2-10^3$ of these integrations. Note that, to keep the memory footprint low, we deliberately do not use the convolution theorem to compute the susceptibilities via FFTs as this would require a much more dense (and uniform) grid with $\sim10^3$ frequency points to resolve all features. Although this still fits into memory for the approximations considered in this work and reduces runtime since it is much more parallelizable, an extension to a full treatment of the vertex would become prohibitively expensive. Note that implementing a compression scheme like the quantics tensor cross interpolation (QTCI)~\cite{Shinaoka_2023, Ritter_2024, Rohshap_2025} might allow to exploit FFTs in a more general setting.

\subsubsection{Numerical parameters}
In order to adequately resolve all features on the real frequency axis, we set the frequency grid resolution to $N_{\omega}=201$ with the scale factor $B=15$. In the FRG scheme, we initialize the flow at $\Lambda_{\mathrm{ini}}=10.0$ with the SCHF dressed RPA solution and then flow to $\Lambda_{\mathrm{fin}}=0.0$. Choosing $\gamma$ then allows tuning $U/\Delta$, $\Omega/\Delta$, $A/\Delta$ at each point during the flow. If not mentioned otherwise, we will set $U=\Omega=A=1.0$. The number of Floquet coefficients kept to reach convergence is $n_f^{(2)} =11 - 21$ for the drive protocols considered. \\
In the self-consistency schemes, we iterate until both filling (see Eq.~\eqref{eq: filling}) and self-energy are converged below a threshold of $10^{-4}$.

\section{Results}\label{section: Results}
We consider the $\mathrm{SU2}$-symmetric SIAM with metallic leads. The drive protocol is a simple sinusoidal drive of the back gate voltage $V_g$, such that the on-site levels oscillate with
\begin{equation}
    \varepsilon_{\sigma}(t) = V_g +  A \cos(\Omega t)\;.
\end{equation}
The chemical potential of the reservoir is set to $\mu=0$ for all runs (unbiased SIAM).
We aim to benchmark the newly developed FRG framework against 2PT and GW, as exact results are not available. 
Additionally, we also compare results for the ``static'' approximations rsFRG and sFRG, whenever static quantities are discussed.
None of the results shown include inter-channel mixing, since we found the static-feedback scheme introduced in Ref.~\cite{Jakobs_2010a} to produce unphysical results and the full parametrization of the two-particle vertex stays out of reach. Finally, we study how many-body correlations affect the Floquet physics by investigating the Kondo resonance and Floquet engineered transport through the dot.

\begin{figure}[htb]
\includegraphics[width=\columnwidth]{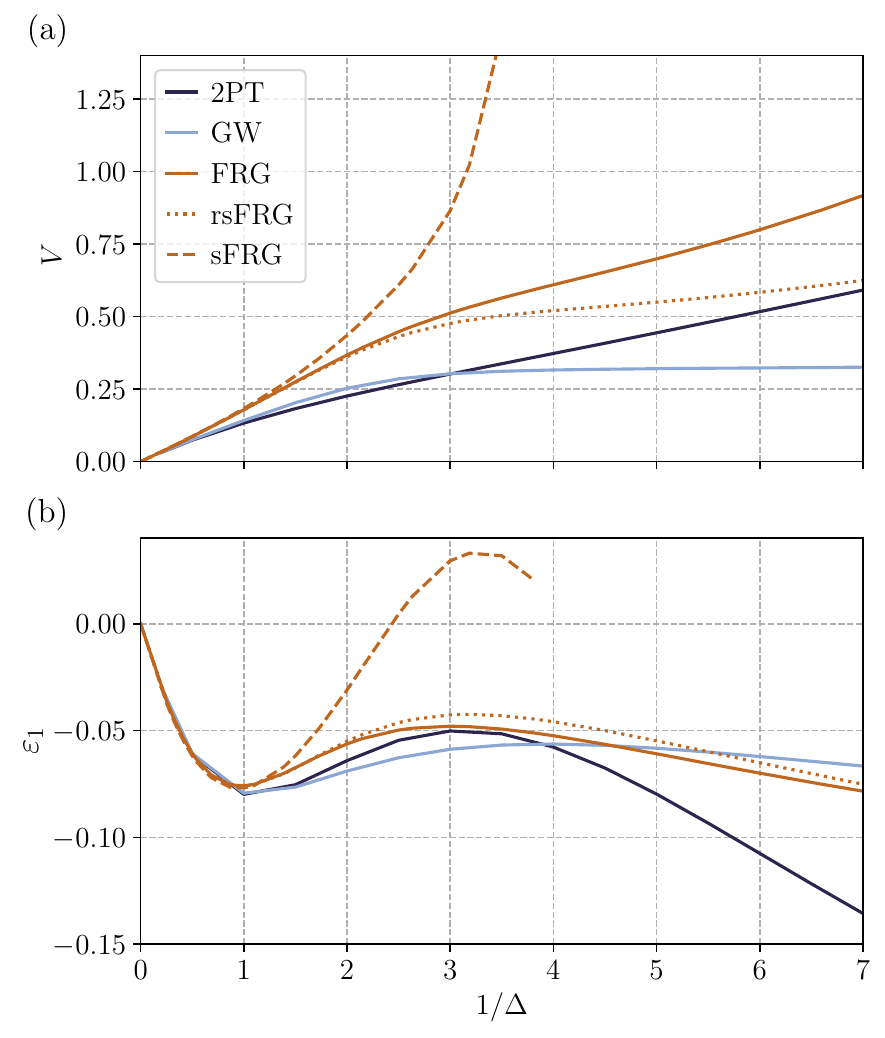}
\centering
\caption{Renormalization of the time-averaged effective interaction $V$ and the $n=1$ component of the dot level $\varepsilon_{1}$ as a function of $1/\Delta$. As comparison, 2PT and GW results are shown. For the FRG, the three different approximations summarized in Fig.~\ref{fig:FRG approximation table} are shown with solid, dashed and dotted lines respectively. The sFRG simulation diverges for $1/\Delta\approx \pi$ as in the equilibrium case.
}
\label{fig:renormalized parameters}
\end{figure}
\subsection{Renormalized interaction and levels} \label{section: renormalized parameters}
Fixing $U=1.0$, we conduct a single $\Delta$-sweep to investigate the renormalization of the system parameters during the flow to large $U/\Delta$.
Fig.~\ref{fig:renormalized parameters}(a) shows the renormalization of the time-averaged effective interaction $V\equiv V_0$ as obtained from Eq.~\eqref{eq: effective interaction} as a function of $1/\Delta$ for the different approaches. For small $1/\Delta$, i.e., small interaction strengths, the static approximations are practically equivalent to the dynamical simulations. For intermediate $1/\Delta \gtrsim1.5$ the sFRG begins to diverge and the flow breaks down at $1/\Delta \approx \pi$. This means that one needs to take finite frequency effects into account to properly renormalize the vertex in the intermediately interacting regime already, similar to what is known from equilibrium Hartree-Fock simulations of the SIAM~\cite{Jakobs_2010a}. However, most of these effects are situated at multiples of $\Omega/2$, as can be seen from the rsFRG results that stay very close to the FRG with fully frequency-dependent quantities. Generally, all FRG approaches show a stronger renormalization than the other approaches, which can be attributed to the bias towards ladder diagrams. \\
We further investigate the performance of the different approaches by considering the renormalization of the level energy on the dot $\varepsilon_n \equiv \mathrm{Re} \Sigma_n^{\mathrm R}(\omega=0)$. Due to the drive protocol, there is no renormalization of the time average, it stays at the particle-hole symmetric position of $U/2$. However, the drive induces oscillations in the level position mediated by the interaction. In the self-consistent schemes, the level renormalization is determined by the sum of the Hartree component and the frequency dependent part, while in the FRG approaches, the level renormalization is part of the self-energy flow. In Fig.~\ref{fig:renormalized parameters}(b) the renormalization of the first harmonic ($n=1$) of the on-site energy is shown. As for the effective interaction, the different approaches only agree in the weakly interacting regime and the static FRG flow breaks down at moderate interaction strength. However, when looking at observables, like occupation number or current (Appendix~\ref{app: real time dynamics}), the differences are very small and only become relevant for large $U/\Delta$ where accuracy is not guaranteed for any of the approaches.\\
In the next sections, we investigate finite-frequency effects specific to the SIAM. As these effects require a dense frequency grid to be appropriately resolved, we do not consider the ``static'' sFRG and rsFRG approximations, the latter only using a grid defined on separated discrete points, which does not represent slopes and curvatures accurately.
\subsection{Self-energy and spectral function}\label{section: spectral}
\begin{figure*}
\includegraphics[width=\linewidth]{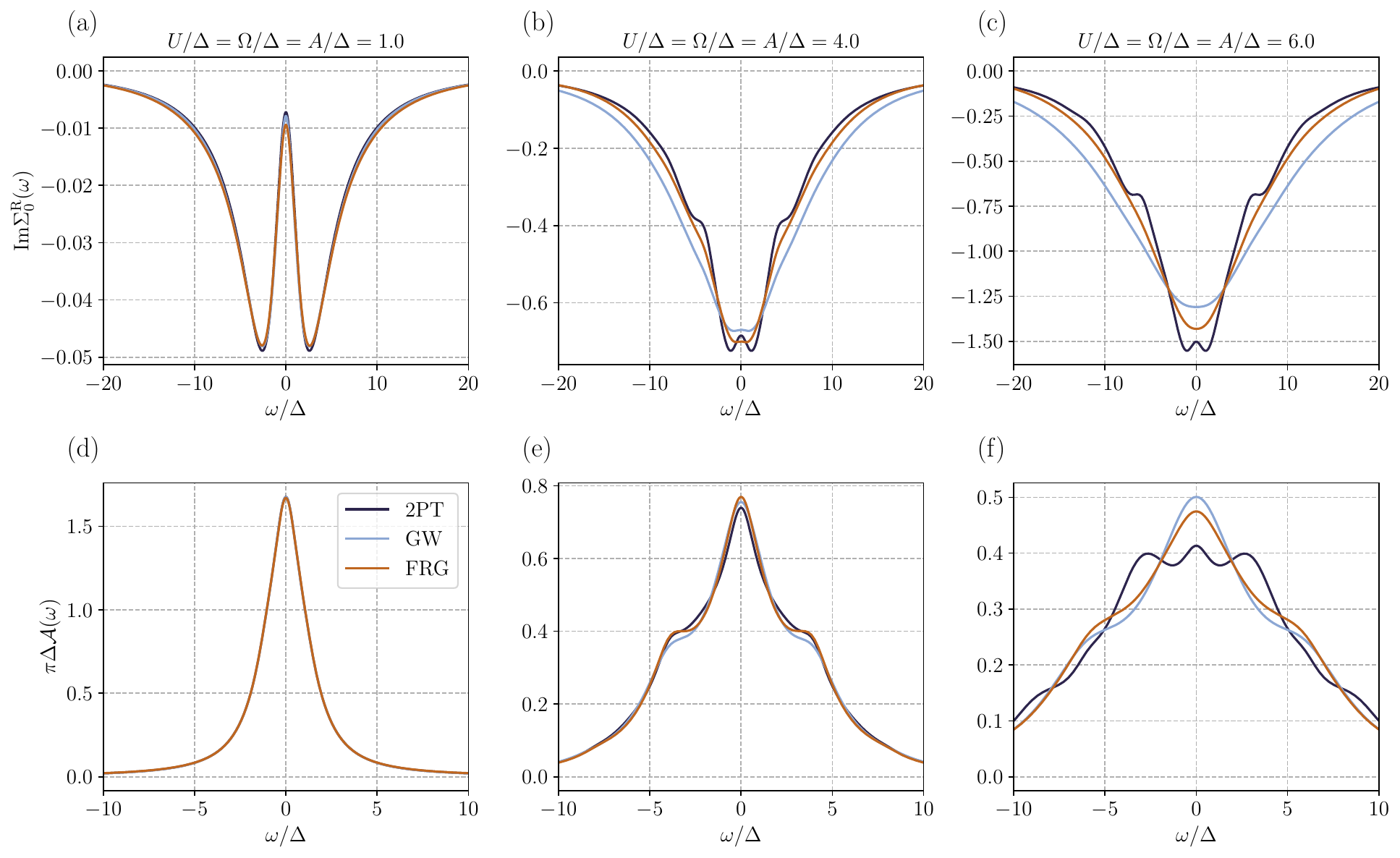}
\centering
\caption{Self-energy and spectral function of the back-gate driven SIAM for different parameters. 2PT, GW, and frequency-dependent FRG results are shown. Panels (a)-(c) show the imaginary part of the time-averaged self-energy. It is negative in all drive-scenarios, i.e., preserves causality. Panels (d)-(f) show the time-averaged spectral function. It is strictly positive for all parameters and can indeed be interpreted as a spectral function. The reservoir temperature is given in units of $\Delta$ as $T_{\mathrm{res}}=0.01\Delta,\; 0.04\Delta,\; 0.06\Delta$ in (a),(d), (b),(e), and (c),(f) respectively.
}
\label{fig:spectral and self-energy}
\end{figure*}
The frequency-dependent simulations (2PT, GW, and FRG) enable the investigation of correlation-induced changes to the spectral function which is modified by the resulting frequency-dependent self-energy.
In Fig.~\ref{fig:spectral and self-energy}(a)-(c) the time-average ($n=0$ component) of the self-energy is shown for a representative set of parameters. At small $U/\Delta$, we find that all methods shown agree well with each other as expected. Even for larger $U/\Delta$ (these are intermediate interaction values from a strong coupling perspective), agreement persists. Importantly, none of the methods shown violates causality [$\mathrm{Im}\Sigma_0(\omega)<0 \;\forall\;\omega$].\\
In the Floquet steady-state, it is generally possible to define a Floquet engineered spectral function that describes the time-averaged particle statistics given by the imaginary part of the Floquet Green's function
\begin{equation}
    \mathcal{A}_n(\omega) = -\frac{1}{\pi}\mathrm{Im} G^{\mathrm R}_n(\omega)\;. \label{eq: spectral function}
\end{equation}
Fig.~\ref{fig:spectral and self-energy}(d)-(f) shows the spectral function for a representative set of parameters. 
In Fig.~\ref{fig:spectral and self-energy}(e), sidebands that correspond to the Floquet replicas appear at $\omega\sim \Omega$. As the sidebands originate from the driven noninteracting model, all approaches capture this. For stronger interaction $U$, the spectral function is significantly modified due to the energetic splitting of single and double occupied dot states. In equilibrium, this is known to lead to the formation of Hubbard bands as well as the Kondo resonance for temperatures below the Kondo temperature $T\lesssim T_K$~\cite{Hewson_1993, Jakobs_2010,Jakobs_2010b,Ge2024} (see also Appendix~\ref{app: equilibrium benchmark}).
 In Fig.~\ref{fig:spectral and self-energy}(f), one can see that both FRG and GW miss the Hubbard splitting at $\omega=\pm U/2$ similar to the equilibrium case. Additionally, all methods predict that the central peak that describes the Kondo resonance is significantly broadened by the drive. 

\subsection{Floquet engineering of the Kondo resonance}\label{section: Kondo}
\begin{figure}
\includegraphics[width=1\columnwidth]{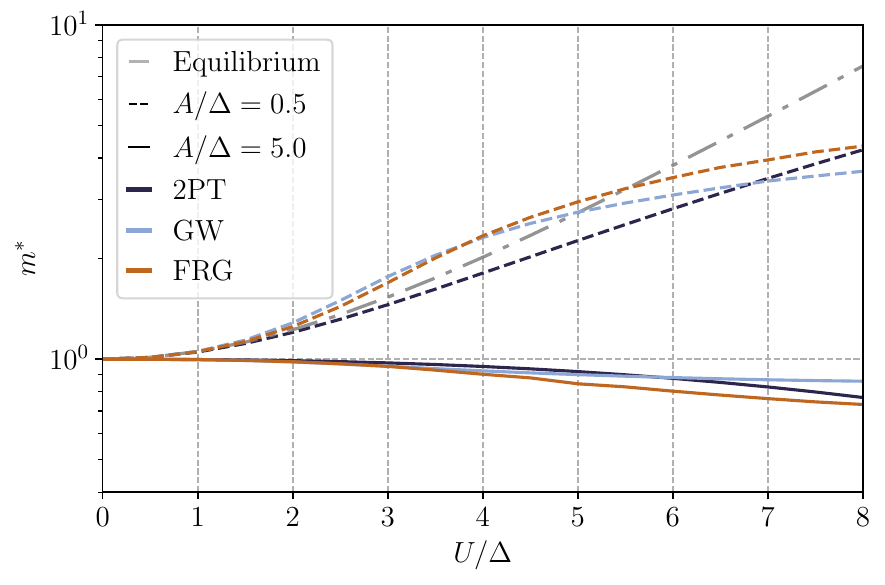}
\centering
\caption{Floquet engineered effective mass by driving the back gate voltage in the SIAM. 2PT, GW and frequency dependent FRG are shown for small ($A/\Delta=0.5$) and large ($A/\Delta=5.0$) drive amplitude. The frequency is set to $\Omega/\Delta=5.0$, which is an intermediate to high drive frequency. The grey dash-dotted line marks the analytically known result for the equilibrium system~\cite{Zlatic_1983}. The reservoir temperature is $T_{\mathrm{res}}=0.05\Delta$, which is still below the Kondo temperature for the shown interaction strengths.}
\label{fig:effective mass}
\end{figure}
\begin{figure*}
\includegraphics[width=\linewidth]{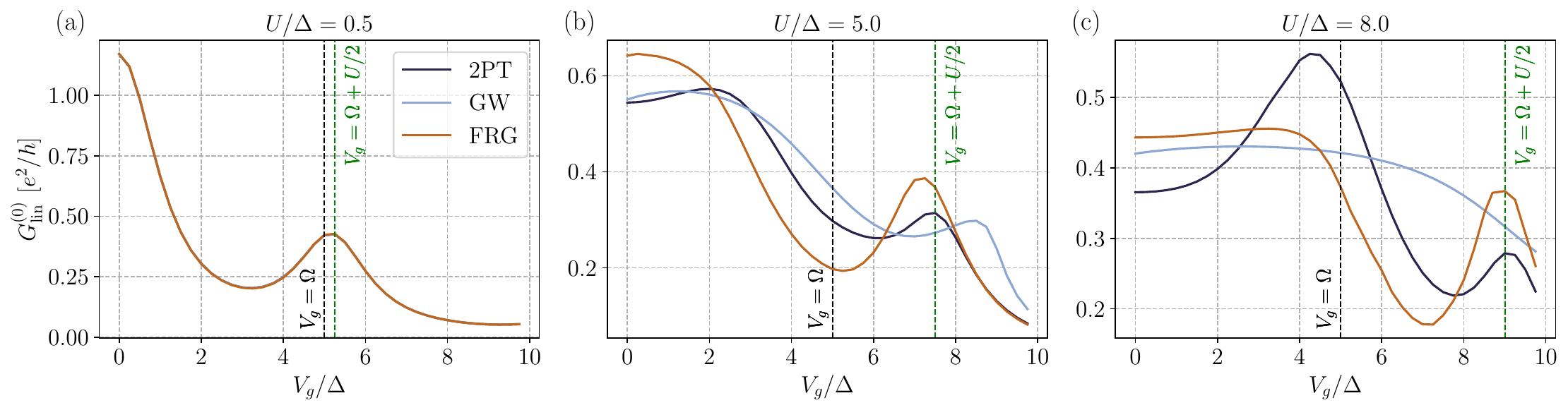}
\centering
\caption{Time averaged linear conductance as a function of gate voltage $V_g$ for different interaction strengths at $A /\Delta = \Omega/\Delta=5.0$. The black dashed line marks the position of the first Floquet replica peak in the noninteracting system at $V_g=\Omega$. The green dashed line marks the position of the Floquet replica of the Hubbard band located at $V_g=\Omega + U/2$. The reservoir temperature is set to $T_{\mathrm{res}}=0.05\Delta$.
}
\label{fig: conductance over gate voltage}
\end{figure*}
The width of the central peak in the spectral function defines the effective mass, $m^* = 1-\partial_{\omega} \mathrm{Re} \Sigma^{\mathrm R}_0(\omega) |_{\omega=0}$, which describes the emergence of the Kondo scale in the strongly correlated limit, where the (equilibrium) SIAM can be mapped to the Kondo model \cite{Hewson_1993, Pustilnik_2004, Andergassen_2008}. Here, the dot is occupied by a single electron that is perfectly screened by the conduction electrons in the reservoirs, leading to a many-body singlet state that is pinned to the Fermi level of the reservoirs.
 As Kondo screening is a nonperturbative many-body effect, it shows exponential behavior $m^*\sim e^{U/\Delta}$, which is known analytically~\cite{Zlatic_1983}. In Fig.~\ref{fig:effective mass}, the effective mass of the driven system is shown as a function of interaction $U$ for $\Omega/\Delta=5.0$. For weak driving ($A/\Delta=0.5$), the effective mass is basically unchanged. The disagreement between the exact equilibrium result and the weak driving results is similar to the disagreement in equilibrium (see Appendix~\ref{app: equilibrium benchmark}).
For strong driving, all methods predict a suppression of $m^*$ that is enhanced with $U$. This can be interpreted as nonequilibrium inelastic scattering suppressing the Kondo resonance, which destroys the coherence of the low-energy dynamical response, in agreement with analytical estimations~\cite{Kaminski_2000}. This result should be checked with a numerically exact method, for example, the one introduced in Ref.~\cite{Sonner_2025}. We expect that the FRG could overestimate the suppression as it does not capture the equilibrium scaling well. Still, the overall agreement of the different methods suggests that this effect is not just an artifact of the FRG. 

\subsection{Correlated photo-induced transport}\label{sec: photo-induced tunneling}
The Floquet sidebands induced by driving the noninteracting system are visible as replica peaks in the spectral function at $\omega=\pm n\Omega$, shown in Fig.~\ref{fig:spectral and self-energy}(e).
Stronger interaction leads to the Hubbard splitting emerging between the Floquet sidebands [Fig.~\ref{fig:spectral and self-energy}(f)] which is, however, only resolved by the bare second order perturbation theory.
In addition to the spectral function, we investigate the time-averaged conductance through the driven dot, since FRG is known to provide good results for the equilibrium conductance, outperforming 2PT~\cite{Jakobs_2010b}.
The Floquet version of the linear conductance follows from Landauer-Bütticker arguments~\cite{Jakobs_2010} as  
\begin{equation}
    G^{(n)}_{\mathrm{lin}} = \frac{e^2}{h}\frac{\pi\Delta}{2} \sum_{\substack{ml \\ \alpha}} \int_{-\Omega/2}^{\Omega/2} \d \omega \; (-\frac{\partial \hat f_{\alpha}}{\partial w})_{m+n,l} \mathcal{\hat A}_{l,m}(\omega)\;,
\end{equation}
with the fermionic reservoir distribution function and the electronic charge $e$. The sidebands in the spectral function transfer to peaks at multiples of $V_g =n\Omega$ in the linear conductance as a function of gate voltage. A photon-dressed hopping process can lead to electron tunneling through the dot despite the dot levels being misaligned with the chemical potential of the leads (which remains fixed at $\mu=0$). 
This is shown in Fig.~\ref{fig: conductance over gate voltage}(a) in the weakly interacting regime, with the first Floquet replica peak visible at $V_g=\Omega$. Stronger interaction [Fig.~\ref{fig: conductance over gate voltage}(b)-(c)] generally reduces the conductance of the driven dot, but it also shifts the replica peak towards larger $V_g$. 
This is because of the Floquet sidebands now decorating the Hubbard bands (double occupied dot) instead of the central Kondo peak. Fig.~\ref{fig: conductance over gate voltage}(b)-(c) show how the peak is, in fact, pushed to $V_g=\Omega+U/2$. This agrees with the results for similar models in the strong coupling regime \cite{Tsuji2008}. 
Strikingly, the GW approximation predicts a totally different behavior where the replica peak disappears entirely. As 2PT and FRG agree at least on the existence and position of the correlation-modified replica peak, we suspect that the GW approximation fails to predict the correct behavior beyond small interaction strengths. \\
For the central peak associated with the singlet state, the results show that, beyond the suppression of the Kondo peak by drive-induced inelastic scattering (see Section~\ref{section: Kondo}), no Floquet sidebands are generated, as such sidebands would otherwise appear at integer multiples of $\Omega$.
 This can be explained by the locality of the drive, which is situated only on the dot, while the singlet state is a non-local many-body state that strongly entangles the dot with the reservoirs (Kondo cloud~\cite{Affleck_2001, Sorensen_2005, Borda_2007}). Thus, the local impurity drive cannot effectively dress a well-defined quasi-particle, and the Floquet sidebands are fully suppressed. Note that this does not contradict results for the (globally) driven Kondo model itself, which allows for sidebands~\cite{Bruch_2022}, as well as Floquet engineering of Kondo channels~\cite{Quito_2023}. 
\\
Finally, the dot-reservoir entanglement of the singlet state is also known to lead to Kondo pinning, which pins the Kondo resonance to the Fermi level of the reservoirs and increases the conductance for small $V_g$ even for larger $U$, as is known from numerically exact simulations~\cite{Jakobs_2010b}.
In Fig.~\ref{fig: conductance over gate voltage}(b)-(c) the 2PT results incorrectly predict a peak in the conductance at $V_g=U/2$ similar to the equilibrium case, missing the Kondo pinning. At equilibrium, the FRG can resolve the correct behavior of the conductance in this regime, and it is reasonable to assume that the same is true for the driven system. The FRG finds that, despite the Kondo resonance itself being suppressed, Kondo pinning partially remains, which shows that the many-body nature of the Kondo cloud is not destroyed by the local drive.

\section{Discussion}\label{section: Discussion}
In this work, we have significantly extended the Floquet–Keldysh functional renormalization group framework by incorporating finite-frequency effects into the flow equations. This represents a substantial improvement over earlier numerical implementations, which were restricted to fully static vertex functions. Benchmarking against established Floquet–Keldysh approaches, including bare second-order perturbation theory and the self-consistent GW approximation, we find excellent agreement in the weak-coupling regime, as expected from perturbative considerations.
We further assessed the validity of a static approximation for the two-particle vertex that only keeps the Floquet coefficients. Our results indicate that this approximation remains quantitatively reliable at weak interaction strengths and thus provides a viable route toward computationally efficient, qualitative studies of driven one- and two-dimensional systems. This opens the possibility of applying Floquet-FRG methods to more realistic models of periodically driven quantum materials.

Finally, we considered correlation effects on Floquet engineered transport through the dot. Driving induces a pronounced peak in the linear conductance through the dot as a function of gate voltage. In the presence of interactions the peak position is shifted from the noninteracting value, $V_g=\Omega$, to $V_g=\Omega+U/2$, indicating that the Floquet sidebands predominantly decorate the Hubbard bands. While the Kondo resonance itself is strongly broadened by drive-induced decoherence, the extended Kondo cloud remains largely unaffected by the local modulation. This is reflected in the complete suppression of Floquet sidebands associated with the Kondo resonance and the partial persistence of Kondo pinning, underscoring the nonlocal and robust nature of the underlying many-body singlet. \\
One can think of several extensions to the present work. On the methodological side, incorporating inter-channel feedback as proposed in Ref.~\cite{Ge2024} is expected to improve the quantitative accuracy at intermediate and strong coupling. This may also address the apparent overestimation of the renormalized interaction strength, potentially in combination with multiloop corrections~\cite{Kugler_2018}. On the application side, a natural next step is to employ the (replica) static approximation to study driven Luttinger liquids in quantum wires~\cite{Andergassen_2004, Klöckner_2020}, or to adopt a truncated-unity FRG scheme~\cite{Profe_2022} for two-dimensional systems, extending recent work in Ref.~\cite{Herre_2026}. The present Floquet-FRG formulation can also be straightforwardly embedded into a Keldysh DMFT solver~\cite{Tsuji2008}. Finally, combining the approach developed here with tensor-network–based compression schemes, such as quantics tensor trains or multiscale space-time ansätze~\cite{Shinaoka_2023, Ritter_2024, Rohshap_2025}, may allow one to retain frequency resolution while achieving sufficient momentum-space resolution for realistic material applications.

\FloatBarrier
\begin{acknowledgements}
We thank Qiyu Liu for fruitful discussions.
This work was funded by the Deutsche Forschungsgemeinschaft (DFG, German
Research Foundation) - 508440990 - 531215165 (Research Unit ``OPTIMAL''). Simulations were performed with computing resources granted by RWTH Aachen University under project rwth1749. 
\end{acknowledgements}
\section*{Data availability}
The data that support the findings of this article are not
publicly available. The data are available from the authors
upon reasonable request.
\begin{appendix}
\section{Equilibrium performance of the FRG} \label{app: equilibrium benchmark}
\begin{figure}[h!]
\includegraphics[width=\columnwidth]{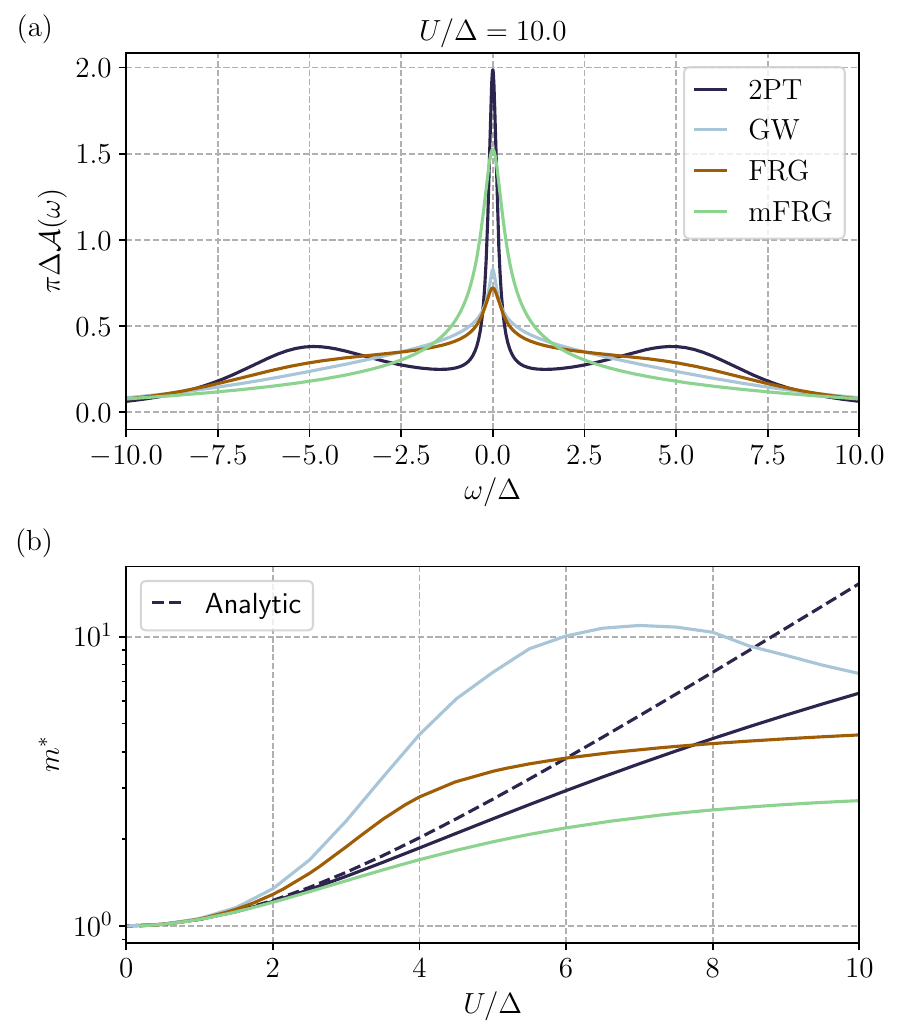}
\centering
\caption{Benchmark of the SIAM at equilibrium, comparing 2PT, GW and different FRG setups. ``FRG'' denotes the frequency-dependent FRG without inter-channel mixing that is used for the main text results. ``mFRG'' denotes frequency-dependent FRG with static inter-channel mixing like in Ref.~\cite{Jakobs_2010b}. (a) Spectral function at intermediate interaction strength, where the Hubbard bands become visible. (b) Effective mass as a function of interaction strength. Dashed grey line marks the analytically known exponential scaling. Note, that compared to Ref.~\cite{Jakobs_2010b} the x-axis must be scaled by a factor $1/2$.}
\label{fig: equilibrium benchmark}
\end{figure}
In this appendix, we provide a performance benchmark of the implemented FRG schemes on the $\mathrm{SU}2$-symmetric SIAM at zero magnetic field, $\mu=0$ and $T=0.01$. The analytical results~\cite{Zlatic_1983} are obtained at $T=0$. For insights on the performance of FRG schemes that treat the full frequency dependence of the two-particle vertex, see Ref.~\cite{Ge2024}. Generally, on a one-loop level, a full treatment of the vertex does not qualitatively improve the accuracy. In Fig.~\ref{fig: equilibrium benchmark}(a) the spectral function is shown for the different approximations considered. 
In Ref.~\cite{Jakobs_2010b} a static self-energy feedback scheme was introduced that was reported to perform better at $T=0$. 
Further, static inter-channel mixing of the vertex was introduced to fix the RPA divergence. It turns out that none of this is necessary or beneficial at finite $T$ in combination with the Katanin substitution. For completeness, we still show a version with static inter-channel mixing (''mFRG'') in Fig.~\ref{fig: equilibrium benchmark}(a). Apart from the 2PT, none of the methods captures the development of the Hubbard bands. The method that performs best, showing at least a slight indication of the Hubbard bands at the correct position, is the FRG, which uses no inter-channel mixing, fully frequency-dependent self-energy feedback and the Katanin substition.\\
For the case of driven systems discussed in the main text, we found additional problems with the approach in Ref.~\cite{Jakobs_2010b}. For instance, the static self-energy feedback leads to unphysical, causality-breaking self-energies with similar effects appearing when using the static inter-channel mixing scheme (mFRG). This is due to the coupling between different Floquet components not being incorporated correctly when using these static feedback mechanisms.
Therefore, we use frequency-dependent feedback with Katanin substitution and without static inter-channel mixing for all simulations except for the fully static approximation ``sFRG'' shown in Fig.~\ref{fig:renormalized parameters}. \\
In Fig.~\ref{fig: equilibrium benchmark}(b), the equilibrium effective mass is shown as a function of $U/\Delta$. All methods fail to capture the exponential scaling that is characteristic of the Kondo effect. The GW and the FRG without channel mixing show an initial overestimation of the effective mass that can be attributed to the bias towards the RPA ladder diagrams. At finite $T$ and with the Katanin substitution the FRG does not break down, in contrast to the case considered in Ref.~\cite{Jakobs_2010b}. The mFRG tends to severely underestimate the effective mass even compared to 2PT. In the driven system, using no inter-channel mixing, an overestimation of the effective mass remains for the case of weak driving (see Fig.~\ref{fig:effective mass}).

\section{Real-time steady-state dynamics} \label{app: real time dynamics}
\begin{figure}
\includegraphics[width=1\columnwidth]{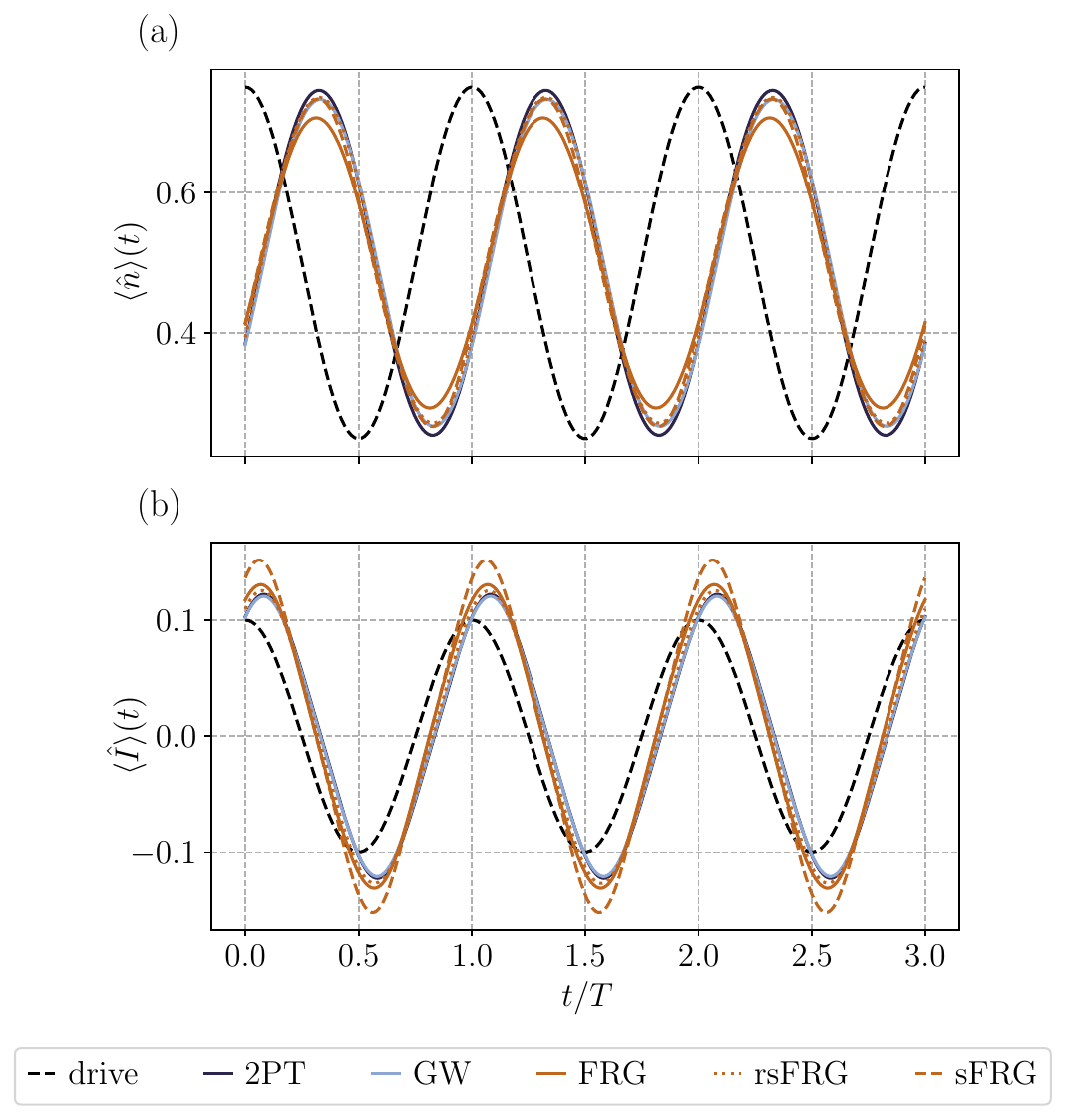}
\centering
\caption{Real-time steady-state dynamics of the dot filling and current through the dot. (a) Dot filling of the (particle-hole symmetric) SIAM with driven back gate voltage obtained with the different approximation introduced in this work. The black dashed line indicates the drive shape ($\cos\Omega t$). (b) Current through the dot in units of $\displaystyle \frac{e}{\hbar}$. The back-gate voltage drive induces an alternating current. The parameters are $U=\Omega=A=2\Delta$, $T_{\mathrm{res}}=0.02\Delta$.}
\label{fig: real-time dynamics}
\end{figure}
\begin{figure*}
\includegraphics[width=\linewidth]{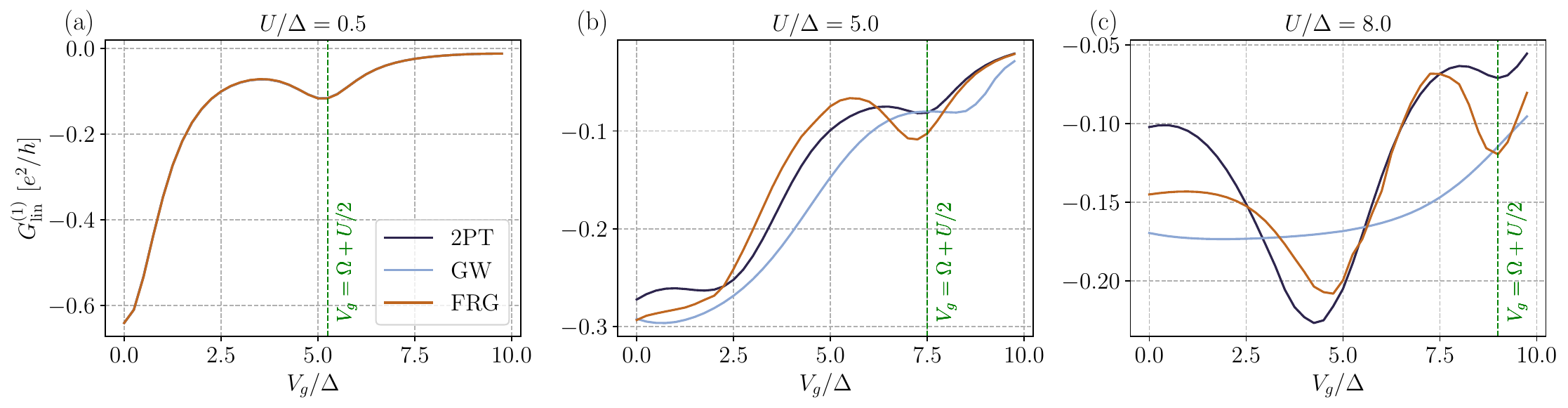}
\centering
\caption{First Floquet component ($n=1$) of the linear conductance as a function of gate voltage $V_g$ for different interaction strengths at $A /\Delta = \Omega/\Delta=5.0$. The green dashed line marks the position of the Floquet replica of the Hubbard band located at $V_g=\Omega + U/2$. The reservoir temperature is set to $T_{\mathrm{res}}=0.05\Delta$.
}
\label{fig: conductance over gate voltage first harmonic}
\end{figure*}
The dot occupation number can be computed similarly to the Hartree self-energy
\begin{equation}
    \langle \hat n\rangle_{n} =  \frac{1}{2\pi i} \int_{-\infty}^{\infty} \d \omega \;G^{<}_n(\omega)\;. \label{eq: filling}
\end{equation}
and the steady-state dynamics of the dot occupation can be recovered by
\begin{equation}
    \langle \hat n\rangle(t) = \sum_n e^{-in\Omega t}\langle \hat n\rangle_n\;. \label{eq: real time}
\end{equation}
Following Eqs.~\eqref{eq: filling} and~\eqref{eq: real time} we can compute the steady-state dynamics of the dot filling which should be synchronized with the drive. This is shown in Fig.~\ref{fig: real-time dynamics}(a).\\
 We can also investigate the alternating current that is induced by the back-gate drive, shown in Fig.~\ref{fig: real-time dynamics}(b). 
The steady-state current through the dot can be computed with the Heisenberg equation of motion. For a given lead $\alpha$, the current through this lead
is obtained through $\langle \hat I\rangle_{\alpha, n} = (\langle \hat j\rangle_{\alpha, n} + \langle \hat j\rangle^{*}_{\alpha, -n}) / 2$ with the Floquet coefficients given in terms of a sum over Floquet matrices
\begin{align}
    \langle \hat j\rangle_{\alpha, n} =   \frac{e}{\hbar}\sum_{ml} \int_{-\Omega/2}^{\Omega/2}\frac{\d \omega}{2\pi} &\; \hat\Sigma_{\alpha}^{\mathrm R}(\omega)_{m+n,l}\hat G^{\mathrm K}(\omega)_{l,m} \nonumber\\ 
    &+ \hat\Sigma_{\alpha}^{\mathrm K}(\omega)_{m+n,l}\hat G^{\mathrm A}(\omega)_{l,m}\;. \label{eq: current}
\end{align}
Both filling and current oscillate in synchronization with the drive up to a phase factor. For the shown (moderate) interaction strength, the different methods agree qualitatively, including the static approximations. 

\section{First harmonic of the linear conductance} \label{app: conductance}
In addition to the time averaged linear conductance shown in Fig.~\ref{fig: conductance over gate voltage} in the main text it is also insightful to investigate the higher harmonics that describe oscillations of the linear conductance. In Fig.~\ref{fig: conductance over gate voltage first harmonic} the first harmonic ($n=1$ Floquet component) of the linear conductance is shown as a function of gate voltage for the same parameters as in Section~\ref{sec: photo-induced tunneling}. The absolute value of the oscillations shows similar behavior to the time average, i.e., at gate voltages where the averaged conductance is enhanced ($V_g=\Omega + U/2$) the oscillation amplitude also increases. We find no additional features in the higher order harmonics that could indicate a higher harmonic response.
\end{appendix}
\FloatBarrier
%

\end{document}